\documentclass[lettersize, journal,twoside]{IEEEtran}

\usepackage{amsmath,amssymb,amsfonts} 
\usepackage{graphicx, scalerel}
\usepackage{textcomp}
\usepackage{comment}

\usepackage[most]{tcolorbox}
 
\usepackage[linesnumbered,ruled]{algorithm2e}
\newtheorem{theorem}{Theorem}
\newtheorem{remark}{Remark}

\usepackage{pgfplots}
\usetikzlibrary{shapes,arrows}
\usepackage{booktabs}
\usepackage{cite}
\usepackage{todonotes}  
\newcommand\sbullet[1][.5]{\mathbin{\ThisStyle{\vcenter{\hbox{%
  \scalebox{#1}{$\SavedStyle\bullet$}}}}}%
}

\definecolor{mst1}{rgb}{0.3266,    0.1840,    0.4290}
\definecolor{mst2}{rgb}{0.9665,    0.6340,   0.1526}
\definecolor{mst3}{rgb}{0.7629,    0.1073,    0.4411}
 
 \newcommand{\hide}[1]{}
 \newcommand{\ovl}{\overline}
 \newcommand{\unl}{\underline}
  \newcommand{\oa}{\ovl{\alpha}}
  \newcommand{\ua}{\unl{\alpha}}
  \newcommand{\da}{{\alpha}^\Delta}
\newcommand{\KK}{\mathcal{K}}
 \newcommand{\KI}{\KK_\infty}

\begin{document}  

\title{Automated Controller Calibration \\ by Kalman Filtering} 
\author{Marcel Menner, %\IEEEmembership{Member, IEEE}, 
Karl Berntorp, %\IEEEmembership{Senior Member, IEEE}, \\
and Stefano Di Cairano %\IEEEmembership{Senior Member, IEEE}
%\thanks{The research in this manuscript was not funded by any government agency.}
%\thanks{Manuscript received 09 August 2022; revised 22 January 2023; accepted 03 March 2023. (Corresponding author: Marcel Menner.)}
\thanks{Marcel Menner, Karl Berntorp, and Stefano Di Cairano are with Mitsubishi Electric Research Laboratories (MERL), 201 Broadway, Cambridge, MA, 02139, USA (e-mail: menner@ieee.org; karl.o.berntorp@ieee.org; dicairano@ieee.org).}
\thanks{This article has supplementary downloadable material available at https://ieeexplore.ieee.org, provided by the authors.}
\thanks{Digital Object Identifier 10.1109/TCST.2023.3254213}
%\thanks{\copyright2023 IEEE. Personal use of this material is permitted. Permission from IEEE must be obtained for all other uses, in any current or future media, including reprinting/republishing this material for advertising or promotional purposes, creating new collective works, for resale or redistribution to servers or lists, or reuse of any copyrighted component of this work in other works.}
}

\maketitle
 
%\thispagestyle{empty}
%\pagestyle{empty}

%%%%%%%%%%%%%%%%%%%%%%%%%%%%%%%%%%%%%%%%%%%%%%%%%%%%%%%%%%%%%%%%%%%%%%%%%%%%%%%%
\begin{abstract}
This paper proposes a method for calibrating control parameters. Examples of such control parameters are gains of PID controllers, weights of a cost function for optimal control, filter coefficients, the sliding surface of a sliding mode controller, or weights of a neural network. Hence, the proposed method can be applied to a wide range of controllers. The method uses a Kalman filter that estimates control parameters, using data of closed-loop system operation. The control parameter calibration is driven by a training objective, which encompasses specifications on the performance of the dynamical system. The performance-driven calibration method tunes the parameters online and robustly, is computationally efficient, has low data storage requirements, and is easy to implement making it appealing for many real-time applications. Simulation results show that the method is able to learn control parameters quickly, is able to tune the parameters to compensate for disturbances, and is robust to noise. A simulation study with the high-fidelity vehicle simulator CarSim shows that the method can calibrate controllers of a complex dynamical system online, which indicates its applicability to a real-world system. We also verify the real-time feasibility on an embedded platform with automotive-grade processors by implementing our method on a dSPACE MicroAutoBox-II rapid prototyping unit.
\end{abstract}


%%%%%%%%%%%%%%%%%%%%%%%%%%%%%%%%%%%%%%%
%%%%%%%%%%%%%%%%%%%%%%%%%%%%%%%%%%%%%%%
%%%%%%%%%%%%%%%%%%%%%%%%%%%%%%%%%%%%%%%
%% SECTION
%%%%%%%%%%%%%%%%%%%%%%%%%%%%%%%%%%%%%%%
%%%%%%%%%%%%%%%%%%%%%%%%%%%%%%%%%%%%%%%
%%%%%%%%%%%%%%%%%%%%%%%%%%%%%%%%%%%%%%%
\section{Introduction}
\IEEEPARstart{C}{ommissioning} a control system or designing a controller of a dynamical system requires considerable manual calibration effort in order to meet certain specifications, which slows down the development process. Furthermore, controller calibration is often done at the production stage of the system, while conditions change over the system’s lifetime. Thus, automating controller calibration and enabling online tuning of controllers is relevant for many applications, as it reduces deployment time and cost, and ensures that the system performance remains high throughout the life of the device.  

Controller calibration often aims at making the closed-loop system operation achieve certain specifications. 
Examples of such specifications include reaching a target state within a certain time, avoiding oscillations, or limiting overshooting of a target state. 
In this paper, we propose a method for the automatic calibration of control parameters  driven by such specifications. 
The method can be applied to controller architectures with a given parametric structure, e.g., it can calibrate gains of a PID controller, cost function weights of an optimal controller, filter coefficients for loop shaping techniques, sliding mode controllers, neural network controllers, etc.
Further, the proposed method considers complex system dynamics for the controller calibration, whereas more traditional controller calibration approaches often focus on linear or linearized systems, see, e.g., literature on PID tuning~\cite{aastrom1993automatic}.

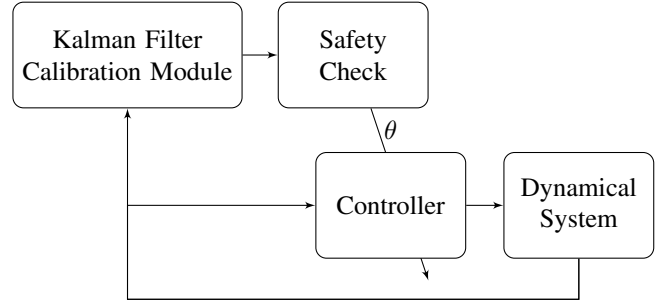
\begin{figure}
    \centering
    %\includegraphics{}
    % Define block styles
    \tikzstyle{decision} = [diamond, draw, fill=blue!20, text width=4.5em, text badly centered, node distance=6cm, inner sep=0pt]
    \tikzstyle{block} = [rectangle, draw, fill=white!20, text width=5em, text centered, rounded corners, minimum height=4em, node distance=3cm]
    \tikzstyle{line} = [draw, -latex']
    \tikzstyle{cloud} = [draw, ellipse,fill=red!20, node distance=3cm, minimum height=2em]
    
    \begin{tikzpicture}[node distance = 2cm, auto]
    \node [block,text width=8em] (tuning) {Kalman Filter Calibration Module};
    \node [block]at(3,0) (safe) {Safety Check};
    \path [line] (safe) --  (4,-3);
    \node [block]at(3.5,-2) (mpc) {Controller};
    \node [block]at(6,-2) (plant) {Dynamical System};
    %\node [below of=plant] (line1) {};
    % Place nodes
    \node []at(3.5,-1) {$\theta$};
    \path [line] (tuning) -- (safe);
    \path [line] (mpc) -- (plant);
    \path [line] (plant) -- (6,-3.25) -- (0,-3.25) -- (tuning);
    \path [line] (plant) -- (6,-3.25) -- (0,-3.25) -- (0,-2) -- (mpc);
    \end{tikzpicture} 
    \caption{Closed-loop tuning of control parameters. The closed-loop of the controller with the dynamical system is augmented by a calibration module that takes data to tune the control parameters, $\theta$, online. The safety check can be used to verify the control parameters for control-theoretic properties such as eigenvalues or Lyapunov stability.}
    \label{fig:block}
\end{figure}

The method is implemented recursively using a Kalman filter that estimates control parameters rather than the state of the dynamical system.
In this setting, the Kalman filter uses a training objective evaluating the performance of the closed-loop system operation online, which is then used to tune the parameters to improve upon the closed-loop system operation measured with respect to the training objective.
The training objective encompasses the specifications for the closed-loop system performance and has a highly flexible structure. 
Fig.~\ref{fig:block} shows the block-diagram of the proposed scheme, where the Kalman filter acts as a tuning module that uses data to calibrate the optimal controller.
The Kalman filter is utilized as a least-squares parameter estimator optimizing a training objective. 
It tracks a joint posterior distribution of the control parameters, which is refined recursively using the training objective and data of closed-loop system operation.

The main advantages of using the proposed Kalman filter are that it
(i) tunes the parameters online during system operation, 
(ii) is robust to noise due to the filter-based design, 
(iii) maintains safety guarantees of the closed-loop operation, 
(iv) is computationally efficient,  
(v) requires reduced data storage due to the recursive implementation, 
and (vi) is easy to implement, hence making it appealing for industrial applications.
In this paper, the method is applied to a state feedback controller, an optimal controller, a PID controller, an $H_\infty$ controller, a sliding mode controller, a dynamic output feedback controller, and a neural network controller, but the method is not restricted to these controllers. Simulation results show that the method achieves fast convergence of the control parameters with a 24\% average decay factor of the closed-loop cost per Kalman filter recursion, improves the closed-loop system performance with a 29\% improvement on tracking precision, and is robust to noise.
Further, we present a simulation study using the high-fidelity vehicle simulator CarSim, in which the Kalman filter method calibrates lane-change controllers. The study shows that the method can calibrate controllers of complex dynamical systems online, which indicates its applicability to a physical system. 

This paper extends our initial study in \cite{menner2021kalman}, where the conceptual idea was introduced for optimal controllers. In this extension, we present the complete study with the generalized framework to include different controller architectures, discuss a safety check that can be used to verify  the control parameters for safety, and verify the method using CarSim and its real-time implementation in embedded architectures on a dSPACE MicroAutoBox-II rapid prototyping unit.

\subsection{Related Work}
The two recent research directions that are most closely related to the method in this paper are Bayesian optimization (e.g., \cite{marco2016automatic,neumann2019data,lu2020mpc}) and retrospective cost optimization (e.g.,~\cite{santillo2010adaptive, Rahman2016, rahman2017retrospective}).
Extended related research directions include iterative learning control (e.g.,~\cite{bristow2006survey, ahn2007iterative, mueller2012iterative, ostafew2013visual}), iterative feedback tuning (e.g.,~\cite{hjalmarsson1998iterative, hjalmarsson2002iterative, hjalmarsson1994convergent, lequin2003iterative, jung2020iterative}), virtual reference feedback tuning (e.g.,~\cite{campi2002virtual, previdi2004data, sala2005extensions, campi2006direct, campestrini2011virtual, care2019toolbox}), and unfalsified control (e.g.,~\cite{safonov1994unfalsified, safonov1997unfalsified, jun1999automatic, bianchi2014fault}), and there is literature that focuses on specific controller classes (e.g., for optimal control \cite{mombaur2010human, clever2016inverse, menner2019constrained, englert2017inverse, chou2020learning, rosbach2019driving, menner2020inverse, hewing2020learning, menner2020maximum, maass2021zeroth}).

\subsubsection*{Bayesian optimization (BO)}
BO-based approaches usually learn a mapping as a black-box function from the control parameters to a pre-specified performance metric, e.g., a reward function, using trial-and-error search~\cite{marco2016automatic,neumann2019data,lu2020mpc}.
Similarly to BO, we use a training objective to learn control parameters.
Differently from BO, we do not require an episodical learning task and we do not have a trial-and-error implementation.
While our method can be applied for episodical learning tasks, it can also be implemented recursively, which allows to compensate for disturbances, while learning the control parameters. 
Furthermore, the recursive implementation is data efficient as the data history need not be stored.

\subsubsection*{Retrospective cost optimization (RCO)}
RCO uses transfer functions and optimizes a cost function retrospectively~\cite{santillo2010adaptive, Rahman2016, rahman2017retrospective}.
RCO adapts certain coefficients such that the new set of coefficients would have led to better performance over a previous window of operation. 
The idea is that the coefficients that would have performed better in the past  will also perform well in the future, which is the case, e.g., in the presence of systematic disturbances.
Similarly to retrospective cost optimization, we can also calibrate an optimal controller retrospectively, which allows for compensating systematic disturbances. 
Differently from retrospective cost optimization, we use a Kalman filter with a training objective to tune generic control parameters beyond optimal control.

\subsubsection*{Iterative Learning Control (ILC)}
Another approach that uses iterations of a repeating task is ILC~\cite{bristow2006survey, ahn2007iterative, mueller2012iterative, ostafew2013visual}. 
ILC does not learn feedback controllers but feedforward control inputs in order to compensate for disturbances and/or unmodeled dynamics, e.g.,~\cite{mueller2012iterative} uses a Kalman filter as a disturbance estimator. 
Hence, ILC relies on repeatability of the specific task, e.g.,~\cite{ostafew2013visual} applies ILC with a disturbance estimator to a mobile robot that repeatedly traverses the same path.
Differently from ILC, the proposed approach does not rely on repetitive tasks, i.e., it can be implemented online for continuous control tasks.

\subsubsection*{Iterative Feedback Tuning (IFT)}
IFT is an approach based on transfer functions and aims to optimize control parameters with respect to a cost function by leveraging gradients~\cite{hjalmarsson1998iterative, hjalmarsson2002iterative}. 
In order to estimate the gradient,~\cite{hjalmarsson1994convergent} shows how to utilize closed-loop experimental data.
IFT is applied to PID tuning in~\cite{lequin2003iterative} with the objective of achieving a fast system response.
In~\cite{jung2020iterative}, IFT is used for tuning both the inner and the outer loop of a cascaded control system.
Similarly to IFT, the proposed approach uses the notion of local improvement of control parameters with respect to a cost function by means of an (approximated) gradient.
Differently from IFT, the proposed Kalman filter implementation provides a framework that recursively updates not only the control parameter values, but also their joint distribution for achieving the best closed-loop performance. 
This joint distribution is essential as it determines how---and how quickly---to adapt the control parameters, which has been observed to be a key benefit in our studies as it improves/accelerates the convergence speed.

\subsubsection*{Virtual Reference Feedback Tuning (VRFT) and Unfalsified Control}
VRFT is an approach that uses data and the transfer function of an ideal closed-loop behavior~\cite{campi2002virtual, previdi2004data, sala2005extensions, campi2006direct, campestrini2011virtual, care2019toolbox}. 
VRFT uses input-output data in order to compute a virtual reference, which is used, in turn, to obtain the control parameters that most closely achieve the ideal closed-loop performance.
Unfalsified control theory uses similar ideas as VRFT of using data of closed-loop system operation~\cite{safonov1994unfalsified, safonov1997unfalsified, jun1999automatic, bianchi2014fault}. 
Differently from VRFT, unfalsified control uses input-output data in order to falsify candidate members of a class of controllers that are not consistent with the performance specifications.  
Some differences from VRFT include that the proposed method is an online-implementable controller calibration method that can incorporate a general selection of training objectives rather than focusing on reference trajectories. 
Differently from unfalsified control, the proposed approach does not use candidate control laws and a falsify mechanism. Instead, it uses data of closed-loop system operation and a Kalman filter to update control parameters and their joint distribution.

\subsubsection*{Extended Literature on Controller Calibration}
Inverse reinforcement learning (IRL) is probably the best known example for calibrating a specific class of controller. IRL uses human demonstrations to learn a cost function and often aims at transferring human expertise to an autonomous system, e.g., for humanoid locomotion~\cite{mombaur2010human, clever2016inverse}, identifying human movements~\cite{menner2019constrained}, robot manipulation tasks~\cite{englert2017inverse,chou2020learning}, or autonomous driving~\cite{rosbach2019driving, menner2020inverse}. 
For model predictive controllers (MPC), a popular research focus lies in designing terminal components to mitigate the potential short-sightedness due to the limited prediction horizon~\cite{brunner2015stabilizing, rosolia2017learning}.
Similarly to IRL, we also learn a cost function.
Differently from IRL, we do not utilize human demonstrations and we do not require an episodical learning task, although our method can be implemented episodically.
Instead, our approach estimates cost function parameters online during system operation. Other examples of methods tailored to specific controller classes include auto-tuning PID controllers~\cite{somefun2021dilemma, kurokawa2018closed, dastjerdi2018tuning}, loop shaping for $H_\infty$ controllers~\cite{glover1992tutorial, mcfarlane1992loop}, and calibrating sliding mode controllers~\cite{edwards1998sliding, young1999control, li1996genetic, hung2007design, mondal2014adaptive, concha2021tuning}.

\subsection{Notation}
\label{sec:prelim}
Given two integer indices $n,m$ with $m\!<\!n$ and a vector $x_i\!\in\!\mathbf{R}^{n_x}$, we define $X_{m|n}\!\in\!\mathbf{R}^{n_x(n\!-\!m\!+\!1)}$ as the vectorized sequence that comprises %the values 
$x_i$ from $i\!=\!m$ through $i\!=\!n$,
\begin{align}
\label{eq:vectorize}
    X_{m|n} & := 
    \begin{bmatrix}
    x_{m}
    \\
    \vdots
    \\
    x_{n}
    \end{bmatrix}.
\end{align}
Further, ${\rm diag}(\lambda)\!\in\!\mathbf{R}^{n_\lambda\times n_\lambda}$ is a matrix, whose diagonal entries are the entries of a vector $\lambda\!\in\!\mathbf{R}^{n_\lambda}$, $I$ is an identity matrix of appropriate dimension, and $0$ is an all-zero matrix of appropriate dimension. 
We define $\|x\|_\Sigma \!:=\!x^T\Sigma x$ and the conditional probability density (PDF) function of a vector $x_k$ at time steps $k=0,...,N$, conditioned on $y$ as $p(x_{0:N}|y):=p(x_0,x_1,...,x_N|y)$.
$\mathcal{N}(\mu,\Sigma)$ is the Gaussian distribution with mean vector $\mu$ and covariance matrix $\Sigma$. 
A continuous function $\alpha\ :\ [0,\ a) \rightarrow [0,\ \infty)$ is said to belong to class $\mathcal{K}$, i.e., $\alpha\in\mathcal{K}$, if $\alpha(0)=0$ and if $\alpha$ is strictly increasing.
Further, $\alpha \in \mathcal{K}_\infty$ if $\alpha\in\mathcal{K}$ and $\alpha(r)\rightarrow\infty$ as $r\rightarrow\infty$.

\subsection{Outline} 
Section~\ref{sec:ProblemStatement} states the problem formulation. 
In Section~\ref{sec:kalman} we present the Kalman filter calibration method used to solve the problem defined in Section~\ref{sec:ProblemStatement}. Section~\ref{sec:doubleInt} contains simulation results for analytical system models, where the Kalman filter is applied to seven different controller architectures and compared to Bayesian optimization. 
Section~\ref{sec:carsim} presents simulation studies, where the Kalman filter is applied to the high-fidelity vehicle simulator CarSim and implemented on a dSPACE MicroAutoBox-II rapid prototyping unit. 
Finally, Section~\ref{sec:conclusion} concludes the paper.
   
%%%%%%%%%%%%%%%%%%%%%%%%%%%%%%%%%%%%%%%
%%%%%%%%%%%%%%%%%%%%%%%%%%%%%%%%%%%%%%%
%%%%%%%%%%%%%%%%%%%%%%%%%%%%%%%%%%%%%%%
%% SECTION
%%%%%%%%%%%%%%%%%%%%%%%%%%%%%%%%%%%%%%%
%%%%%%%%%%%%%%%%%%%%%%%%%%%%%%%%%%%%%%%
%%%%%%%%%%%%%%%%%%%%%%%%%%%%%%%%%%%%%%%
\section{Problem Statement}
\label{sec:ProblemStatement}
We consider discrete-time systems of the form
\begin{align}
\label{eq:sys_dyn}
    x_{k+1} = f(x_k,u_k)+w_k,
\end{align}
where $x_k\!\in\!\mathbf{R}^{n_x}$ is the state at time step $k$, $u_k\!\in\!\mathbf{R}^{n_u}$ is the input, $w_k$ is the process noise or the model mismatch compared to the real-world system, and $f$ is a general nonlinear function. 
The inputs are set by a controller that takes measurements (or estimates) of the state,
\begin{align}
\label{eq:control_law}
    u_k = \kappa_\theta(x_k,z_k),
\end{align}
where $\kappa_\theta$ is the function representing the control law, which could be implicit, e.g., resulting from solving an optimization problem, a pre-determined function, multiplication with gains, or a state-feedback controller.
Here, $z_k$ denotes potentially internal/latent variables such as integral controller variables, filter states, or state estimates.
In this work, we present an approach to calibrate the parameters of the controller, $\theta\in\mathbf{R}^{n_\theta}$, for any controller that is expressed using a parametric structure according to~\eqref{eq:control_law}.
For example, $\theta$ can represent the gains of a PID controller, the weights of a cost function for MPC, or coefficients of a neural network.
To illustrate the applicability of the proposed approach, Section~\ref{sec:doubleInt} applies the method to a selection of seven different controllers.

\subsection{Specifications and Goal of the Calibration Method}
Typically, a control system aims at manipulating the behavior of a dynamical systems such that certain specifications are achieved, in full or as closely as possible. 
Examples of such specifications include minimizing the deviation to a target state, avoiding oscillations, limiting overshooting of a target state, and exceeding a nominal range of operation for not more than a maximum time. 
Our proposed method calibrates the control parameters, $\theta$ in \eqref{eq:control_law}, based on sensor measurements, $x_k$ and $u_k$, and a training objective, which evaluates the performance of the controller and encompasses the specifications for the closed-loop operation of the dynamical system.

The method in this paper adjusts the control parameters online and recursively with a sliding window of data such that the closed-loop system operation maintains its stability guarantees.
As the method uses time-varying parameters, i.e., the parameters are adjusted at each time step, we use time-indices similar to the state and input variables with
\begin{align}
\label{eq:learning_law}
    \theta_{k+1} &=\theta_k + \Delta \theta_k,
\end{align}
where $\theta_k$ are the parameters at time step $k$ that are used in the controller \eqref{eq:control_law}.
The goal is thus to find an adaptation law for the control parameters, $\Delta \theta_k$, such that $\theta_k$ eventually satisfy or optimize certain specifications provided by the system designer, which is discussed next. 

\subsection{Procedure and Training Objective}
The calibration method takes sensor measurements of the state, evaluates the performance of the controller with respect to the specifications, and outputs a new set of parameters for the controller, $\theta_{k+1}$.  
The method calibrates the control parameter vector ${\theta}$ of the generic controller in~\eqref{eq:control_law} to minimize the training objective 
\begin{align}
\label{eq:trainingObj}
    \|{y}_{k} - r(X_{k\!-\!N|k},U_{k\!-\!N|k\!-\!1})\|_{{\rm {C}}_v^{-1}} 
\end{align}
with a positive definite ${\rm {C}}_v$, desired nominal values ${y}_{k}$, and specification function,
\begin{equation}
\label{eq:eval}
r(X_{k\!-\!N|k},U_{k\!-\!N|k\!-\!1})\in\mathbf{R}^{n_r}.
\end{equation} 
Hence, the specification function~\eqref{eq:eval} takes the past state and input sequences (of length $N$), $X_{k\!-\!N|k}$ and $U_{k\!-\!N|k\!-\!1}$, and compares the computed specifications in~\eqref{eq:eval} with a nominal value $y_k\in\mathbf{R}^{n_r}$ that the controller should ideally achieve.
One difference between the training objective and a cost function for optimal control is that the training objective is generic in its structure and flexible in its specifications. 
For example, the training objective can include non-analytic functions.
On the other hand, the cost function often requires a certain structure that enables real-time implementation, e.g., suited for convex optimization. 

\begin{remark}
Our approach is related to Bayesian optimization, where a reward/loss is obtained after a trial. 
The main conceptual difference is the online capable, recursive, and real-time feasible implementation, which uses a Kalman filter-based design outlined in Section~\ref{sec:kalman} rather than a trial-and-error search. 
\end{remark} 
\begin{remark}
The method proposed in this paper is similarly applicable if only a subset of the parameters are to be adjusted with $\theta_{k+1}=\theta_k+G\Delta\theta_k$, e.g., if certain parameters are known or should be fixed.
For ease of exposition, however, we use \eqref{eq:learning_law} throughout. 
\end{remark}

%%%%%%%%%%%%%%%%%%%%%%%%%%%%%%%%%%%%%%%
%%%%%%%%%%%%%%%%%%%%%%%%%%%%%%%%%%%%%%%
%%%%%%%%%%%%%%%%%%%%%%%%%%%%%%%%%%%%%%%
%% SECTION
%%%%%%%%%%%%%%%%%%%%%%%%%%%%%%%%%%%%%%%
%%%%%%%%%%%%%%%%%%%%%%%%%%%%%%%%%%%%%%%
%%%%%%%%%%%%%%%%%%%%%%%%%%%%%%%%%%%%%%%
\section{Kalman Filter for Control Parameter Calibration}  
\label{sec:kalman}
In this section, we propose the recursive algorithm to calibrate or adjust the parameters of a generic controller.  
Given the parameters at the current time step, $\theta_k$, we  perform an update, $\Delta \theta_k$, based on data of closed-loop system operation as in~\eqref{eq:learning_law} aiming at minimizing the training objective~\eqref{eq:trainingObj}.
Hence, the learning method makes 
\eqref{eq:eval} follow a nominal value $y_k$ ``as closely as possible", i.e., the dynamical system is controlled perfectly if $y_k\!=\!r(X_{k\!-\!N|k},U_{k\!-\!N|k\!-\!1})$.
In order to ease exposition, using the notation in~\eqref{eq:vectorize}, we define a, possibly nonlinear, function $F$ with
\begin{align}
\label{eq:sys_dyn_stacked} 
    X_{k|k+N+1}
    &=
    F(U_{k|k+N},x_k,W_{k|k+N}),
\end{align}
where $W_{k|k+N}$ denotes the stacking of $w_k$ in~\eqref{eq:sys_dyn} from time $k$ to $k+N$, i.e., \eqref{eq:sys_dyn_stacked} is equivalent to applying~\eqref{eq:sys_dyn} iteratively from $k$ to $k+N$.
Let $y_k\! =\! r(X_{k\!-\!N|k},U_{k\!-\!N|k\!-\!1})\! +\! v_k$, where $v_k$ denotes a slack variable. 
As the input sequence $U_{k\!-\!N|k\!-\!1}$ results from the state sequence due to feedback control as in~\eqref{eq:control_law} that is fully determined by the initial state of the dynamical system at time $k-N$, $x_{k\!-\!N\!}$, the initial state of the controller (if the controller has internal states), $z_{k\!-\!N\!}$, the model mismatch $W_{k\!-\!N\!|k\!-\!1}$, as well as the control parameters, $\theta_k$.
Therefore, the evaluation function, $r$, can be replaced by a function $r_0$ with
\begin{align*}
    y_k  
    & = r(X_{k\!-\!N|k},U_{k\!-\!N|k\!-\!1})+ v_k
    \\
    & = r_0(\theta_k,x_{k\!-\!N\!},z_{k\!-\!N\!},W_{k\!-\!N\!|k\!-\!1}) + v_k.
\end{align*}
Note that $W_{k\!-\!N\!|k\!-\!1}$ can readily be computed with \eqref{eq:sys_dyn} using state measurements and inputs.
To further ease notation, we define
$$h(\theta_k):=r_0(\theta_k,x_{k\!-\!N\!},z_{k\!-\!N\!},W_{k\!-\!N\!|k\!-\!1}).$$ 
Finally, we obtain the following two main equations for the tuning method proposed in this paper,
\begin{subequations}
\begin{align}
     \theta_{k+1} &=\theta_k + \Delta \theta_k
    \\
    \label{eq:measurement_equation}
     y_k &= h(\theta_k) + v_k.
\end{align}
\end{subequations}
In order to drive the adaptation law, we model the parameters, $\Delta\theta_k$, as well as the slack variable, $v_k$, as i.i.d. random variables with Gaussian zero-mean prior distributions 
\begin{equation}\label{eq:noise}
\Delta \theta_k^{\rm prior}\!\sim\! \mathcal N(0,{\rm C}_\theta) \text{ and }v_k\! \sim\! \mathcal {N}(0,{\rm C}_v).
\end{equation}
As a result, we obtain the parameter adaptation law from the corresponding posterior distribution \begin{equation}\label{eq:posterior}
    p(\theta_{k+1}|\theta_{0:k},y_{0:k})=\prod_{i=0}^{k} p(\theta_{i+1}|\theta_{i},y_{i})p(\theta_0),
\end{equation}
estimated by a Kalman filter where $y_k$ is a vector of desired values for the system operation rather than sensor measurements.
One benefit of using the Kalman filter is that it allows for a recursive implementation, 
which means that we do not need to store the entire data history but only the data used 
for the Kalman-filter recursion. 
In \eqref{eq:measurement_equation}, $h(\theta_k)$ considers process noise (or the model mismatch), $W_{k\!-\!N|k\!-\!1}$, explicitly. Hence, we seek to find the parameters $\theta_k$ that optimize the training objective for the process noise distribution/disturbances at hand. 
This allows for accommodating systematic disturbances as described in Section~\ref{sec:doubleInt}.
In the following, we present two implementations of the Kalman filter parameter tuning method and discuss stability properties.

\subsection{Extended Kalman Filter}
If $h(\theta_k)$ is differentiable with respect to the control parameters $\theta_k$, the following implementation of the extended Kalman filter (EKF) can be used to compute the parameter update in \eqref{eq:learning_law},
\begin{subequations} 
\label{eq:EKF}
\begin{align} 
    \Delta \theta_k
    =
    K_k \left( y_k - h(\theta_k) \right)
\end{align}
with Kalman gain $K_k$ computed as 
\begin{align}
    K_k
    &=
    P_{k|k-1} H_k^T S_k^{-1},
    \\
    S_k 
    &=  
    H_k P_{k|k-1} H_k^T + {\rm C}_v ,
    \\
    P_{k|k-1} 
    &=
    P_{k-1|k-1} + {\rm C}_\theta ,
    \\
    P_{k|k} 
    &=
    \left(
    I - K_k H_k
    \right)
    P_{k|k-1}, 
\end{align}
\end{subequations}
where $H_k=\frac{\partial}{\partial \theta}h(\theta)\ |_{\theta=\theta_k}$, $S_k$ is the innovation covariance, and $P_{k|k}$ is the estimate covariance matrix.
In order to compute the linearization $H_k$, we use the chain rule  
\begin{align*}
    H_k = \frac{\partial h(\theta)}{\partial \theta} 
    = 
    \frac{\partial h(\theta)}{\partial z} \frac{\partial z}{\partial \theta}  
    ,\quad
    z = \begin{bmatrix}
X_{k-N|k} \\ U_{k-N|k-1}
\end{bmatrix}.
\end{align*}
The only step of the EKF implementation that is (potentially) computationally demanding for some controller configuration is the gradient computation $H_k$. 
For optimization-based controllers, we can use $\frac{\partial z}{\partial \theta}\approx \frac{\Delta z}{\Delta \theta}$, where $\frac{\Delta z}{\Delta \theta}$ is obtained by differentiating the KKT conditions.  
This strategy to compute sensitivities is based on applying the Implicit Function Theorem to the KKT conditions~\cite{Fiacco:1253810,pirnay2012optimal}.

\begin{remark}
We use the EKF implementation to refer to a gradient-based method.  
However, as outlined in \cite{Gustafsson2012}, there exist implementations of an EKF that avoid computing the gradient explicitly. 
\end{remark}    

\subsection{Unscented Kalman Filter}
\label{ssec:UKF}
An even more promising and attractive alternative to the EKF is to use an unscented Kalman filter (UKF) to compute the Kalman gain $K_k$.
The UKF uses deterministic samples (called sigma points) around the mean, which are propagated and used to update the mean and covariance estimates~\cite{terejanu2011unscented}. The sigma points are computed using the current best estimate of the distribution of the control parameters, defined by the mean and the covariance.
In the following, we use super-scripts, ${\rm sp}i$, to index sigma points, as opposed to the subscripts indicating the time step, $k$.
Using a UKF, $h(\theta_k)$ need not be differentiable, which is one major advantage of this implementation.
Here, the adaptation law for the parameters is given as
\begin{subequations}
\label{eq:UKF}
\begin{align}
 \Delta \theta_k = K_k (y_k - \hat y_k)   
\end{align}
with the Kalman gain $K_k$ computed using evaluations of sigma points with respect to the specification function~\eqref{eq:eval},
\begin{align}
    \hat \theta_k
    & = 
    \textstyle
    \sum_{i=0}^{2L}w^{a,i} \theta^{{\rm sp}i}_k
    \\
    \label{eq:sigmapoints_h}
    y^{{\rm sp}i}_k
    & =
    h(\theta^{{\rm sp}i}_k)
    \\
    \hat y_k
    & = 
    \textstyle
    \sum_{i=0}^{2L}w^{a,i} y^{{\rm sp}i}_k
    \\
    K_k
    &=
    C_{sz}S_k^{-1}
    \\
    S_k 
    &=  
    \textstyle
    {\rm C}_v + \sum_{i=0}^{2L}w^{c,i}(y^{{\rm sp}i}_k - \hat y_k)(y^{{\rm sp}i}_k -\hat y_k)^T
    \\
    C_{sz}
    &=
    \textstyle
    \sum_{i=0}^{2L}w^{c,i}(\theta^{{\rm sp}i}_k-\hat \theta_k)(y^{{\rm sp}i}_k - \hat y_k)^T
    \\
    P_{k|k-1} 
    &=
    \textstyle
    {\rm C}_\theta + \sum_{i=0}^{2L}w^{c,i}(\theta^{{\rm sp}i}_k-\hat \theta_k)(\theta^{{\rm sp}i}_k-\hat \theta_k)^T
    \\
    P_{k|k} 
    &=
    P_{k|k-1} 
    - K_k S_k K_k^T,
\end{align}
\end{subequations}
where $\theta^{{\rm sp}i}_k$ with $i\!=\!0,...,2L$ are the sigma points, $w^{c,i}$ and $w^{a,i}$ are the weights of the sigma points, $C_{sz}$ is the cross-covariance matrix, $S_k$ is the innovation covariance, and $P_{k|k}$ is the estimate covariance.

The main advantage of the UKF-based implementation is the ability to embed non-differentiable objectives, as gradients do not need to be computed. Instead, evolutions of the system dynamics are simulated for all $2n_\theta\!+\!1$ sigma points, where $n_\theta$ is the number of control parameters, which are then used to obtain the update direction with the Kalman gain as in~\eqref{eq:UKF}.
This advantage is also highlighted in~\cite{menner2022learning}, where digital twin simulations are used to obtain a control policy by means of a UKF-based solver.
In this paper, we choose the following weights and computation of sigma points
\begin{align*}
    \theta^{{\rm sp}0}_k 
    &= 
    \theta_k
    \\
    \theta^{{\rm sp}i}_k
    &=
    \theta_k + \sqrt{L/(1-w^0)} 
    [A]^i
    \quad i=1,...,L
    \\
    \theta^{{\rm sp}i}_k
    &=
    \theta_k - \sqrt{L/(1-w^0)}
    [A]^i
    \quad i=L+1,...,2L
\end{align*}
with weights $-1\!<\!w^{0}\!=\!w^{a,0}\!=\!w^{c,0}\!<\!1$, $w^{a,i}\!=\!w^{c,i}\!=\!(1\!-\!w^{a,0})\!/\!(2L)$ and $[A]^i$ being the $i$th column of $A$ with $P_{k\!-\!1|k\!-\!1}=AA^T$, i.e., $A$ is calculated using the Cholesky decomposition.
Other choices of sigma points and weights are also possible.

\begin{remark}[State-Dependent Parameters]
It is straightforward to include a state dependency for the control parameters, e.g.,
\begin{align*}
    u_k = \kappa_{\theta(x_k)}(x_k,z_k).
\end{align*}
This might be useful for applications where the controller is adjusted based on the state, which is often done, e.g., through gain scheduling.
For example, the $i$-th control parameter could be the result of basis functions, $\phi(x)$, and/or depend on certain regions,
\begin{equation}
\label{eq:basis_functions}
\begin{aligned}
    \theta_i(x)=
    \begin{cases}
    \chi_{1,i}^{T}\phi(x) 
    &
    {\rm if}\ \psi(x)\leq \psi_1
    \\
    \chi_{2,i}^{T}\phi(x) 
    &
    {\rm if}\ \psi_1\leq\psi(x)\leq \psi_2
    \\ 
    \chi_{3,i}^T\phi(x) 
    &
    {\rm otherwise}.
    \end{cases}
\end{aligned}
\end{equation}
For \eqref{eq:basis_functions}, the Kalman filter  calibrates  $\theta_i$ indirectly through the weights of the basis functions, $\chi_{1,i}$, $\chi_{2,i}$, and $\chi_{3,i}$, as well as the boundary of the regions, $\psi_1$ and $\psi_2$. 
\end{remark}

\begin{remark}
The UKF implementation is purely based on evaluating sigma points and no gradient computation is needed. Due to the simplicity of implementation and flexibility, we employ the UKF for most of the validation studies in this paper.
\end{remark} 
\begin{remark}
The entries in the training objective may be as simple as the state trajectory with $y_k$ as the reference trajectory.
However, the entries may be any objective that can be calculated from the state and input trajectories. 
For the UKF implementation, the training objective can also be more complex, e.g., to limit the overshoot or oscillations, to avoid exceeding a nominal operating range for longer than given duration, or to achieve an objective within a certain amount of time. The UKF implementation can also include logic in the form of if/then/else statements.
\end{remark}

\subsection{Interpretation of Covariance Matrices}
The interpretation of covariance matrices in the presented framework is different than the interpretation for a classical implementation of the Kalman filter. 
First, the covariance matrix ${\rm C}_v$ defines the relative importance of the components of the vector-valued specification function in~\eqref{eq:eval}. 
In the proposed method, the Kalman filter is utilized as error-driven estimator, where the error is defined by means of the training objective in~\eqref{eq:trainingObj}.
Hence, the Kalman filter drives the control parameter adaptation to minimize the error in~\eqref{eq:trainingObj}. In the classical Kalman filter setting, this error is defined by the measurement error. 
The choice of ${\rm C}_v$ comes natural from the training objective in~\eqref{eq:trainingObj}.
Second, the covariance matrix ${\rm C}_\theta$ determines the aggressiveness of the controller calibration adjustment.
For example, let ${\rm C}_\theta = c_{\rm scale}\cdot I$ with $c_{\rm scale}>0$. 
Then, a smaller $c_{\rm scale}$ makes the algorithm more conservative and potentially converge not as quickly, which may increase robustness.
A higher $c_{\rm scale}$ makes the algorithm more aggressive and potentially converge faster.
For gradient-based methodologies, $c_{\rm scale}$ can be thought of as a (prior) step-size. 
Note, however, that the adaptation rate changes during operation as the joint posterior distribution of the control parameters is updated.
Hence, one advantage of the Kalman filter-based implementation is that this joint distribution defines the spread of the control parameters as they relate to the training objective in~\eqref{eq:trainingObj}.
This spread is leveraged in order to determine how quickly the control parameters can be adjusted, which has been observed empirically to improve convergence.  
For example, for the UKF implementation, this joint distribution is used to compute the sigma points, which define what control parameter realizations should be evaluated.
It is also possible to tune the covariance matrices during application, see the work in~\cite{ardeshiri2015approximate} in which covariance matrices for a Kalman filter are calibrated.

\subsection{Safety Check to Enforce Stability Properties}
The proposed approach for calibrating controller parameters can be augmented with a \emph{Safety Check} module for certifying or enforcing closed-loop stability, as shown in Fig.~\ref{fig:block}. 
The safety check module must ensure two conditions: 
$(i)$ each set of parameters provided to the controller must result in asymptotic stability (AS) of the closed-loop if such parameters are held indefinitely; $(ii)$ the control parameter update should not cause loss of closed-loop AS. According to 
Fig.~\ref{fig:block}, the safety check only affects the output of the Kalman filter-based 
adaptation. Thus, even if the safety check module rejects the updated parameters due to 
violating the  safety conditions, the Kalman filter state is not affected and hence the entire 
data history will be reflected in any future updates of the control parameters. 

Condition $(i)$ involves evaluating a closed-loop with fixed parameters, and hence may 
be verified using any standard method, such as  
eigenvalues, Nyquist arguments, or gain/phase margins, for linear systems. 
Equivalent methods can be applied for the cases of nonlinear controllers, albeit usually 
more complex and applicable only to specific controller classes, e.g.,~\cite{fazlyab2019probabilistic, hu2020reach}~use reachability arguments for neural
network controllers. The ranges 
of parameter values for satisfying these conditions may be also determined offline.  

Condition $(ii)$ can be approached using tools from switched systems 
literature~\cite{liberzon1999stability}. An approach is to assume knowledge of Lyapunov 
functions for the controllers with fixed parameters, which is guaranteed to exist by 
condition $(i)$, and then to show that the combination of such Lyapunov functions 
obtained guarantees the existence of a Lyapunov function for the closed-loop with 
changing parameters, when this is enabled by as appropriate safety check. 
	
Let $\Theta$ be the, possibly unbounded,  set of parameter vectors, and for all 
$\theta\in\Theta$, let $V_\theta(x)$ be 
a Lyapunov function for the closed-loop system when the parameter vector is 
$\theta\in\Theta$, constantly, i.e., there exists 
$\oa_\theta,\ua_\theta,\da_\theta\in\mathcal{K}_\infty$,
such that
\begin{subequations}\label{eq:constantLF}
	\begin{align}
		\ua_\theta(\|x\|) \leq	 V_\theta(x)&\leq \oa_\theta(\|x\|) \\
		 \Delta V_\theta(x)&\leq -\da_\theta(\|x\|),\
	\end{align}
\end{subequations}
where $\Delta V_\theta(x) = V_\theta(f_\theta(x))-V_\theta(x)$, and $f_\theta$ denotes the closed-loop 
system obtained for parameter vector $\theta$. Further, let
\footnote{This assumption is reasonable since the 
	pointwise minimum and maximum of class-$\KI$ functions result  in class-$\KI$ 
	functions.
} $\ua,\oa,\da\in\KI$ where, for all $x$,
\begin{subequations}
\label{eq:Kbounds}
	\begin{align}
\ua(\|x\|)&=\inf_{\theta\in\Theta}\ (\ua_\theta(\|x\|))
\label{eq:Kbounds-inf}
\\ 
\oa(\|x\|)&=\sup_{\theta\in\Theta}\ (\oa_\theta(\|x\|))
\label{eq:Kbounds-sup}
\\
\da(\|x\|)&=\inf_{\theta\in\Theta}\ (\da_\theta(\|x\|))
\label{eq:Kbounds-infd}.
 	\end{align}
\end{subequations}

\begin{theorem}\label{th:AS}
Let~\eqref{eq:constantLF}, \eqref{eq:Kbounds} hold for all $\theta\in\Theta$, and the 
safety check allows the parameter to be updated from $\theta_k$ to $\theta_{k+1}$ at state $x$
only if 
\begin{equation}\label{eq:safetyCheck}
	V_{\theta_{k+1}}(x)\leq V_{\theta_k}(x).
\end{equation}
Then, for the closed-loop system with the control parameter~$\theta$ being 
updated according to~\eqref{eq:EKF} or \eqref{eq:UKF}, and the condition in~\eqref{eq:safetyCheck}, there 
exists 
a Lyapunov function for the state $x$.
\end{theorem}

\begin{IEEEproof}
The sequence of Lyapunov functions $V_{\theta_k}(x_k)$ when the parameter vector 
$\theta$ is time varying, can be seen as function of both state and parameter 
$V(\theta_k,x_k)$. Next, we prove that $V(\theta_k,x_k)$ is a Lyapunov function for the state, $x_k$. 
Consider $\theta_k$, $x_k$, and $\theta_{k+1}$, 
$x_{k+1}$. Due to $V_{\theta_{k}}$ being a Lyapunov function for the closed-loop when 
$\theta_k$ is held constant, we have $V_{\theta_{k}}(x_{k+1})\leq 
V_{\theta_{k}}(x_{k})-\da_{\theta_k}(x_k)$. Due to the condition~\eqref{eq:safetyCheck},
$V_{\theta_{k+1}}(x_{k+1})\leq V_{\theta_{k}}(x_{k+1})$, regardless of whether 
$\theta_{k+1}=\theta_{k}$ or not. Thus,	combining these with~\eqref{eq:Kbounds-infd},	
$$ 
V(\theta_{k+1},x_{k+1})\leq  V(\theta_{k},x_{k}) - \da(\|x_k\|). 
$$ 
Considering~\eqref{eq:Kbounds-inf} and \eqref{eq:Kbounds-sup} providing uniform $\KI$ function bounds with respect to $\theta$, we obtain
\begin{subequations}\label{eq:fullLF}
	\begin{align}
		\ua(\|x_k\|) \leq	 V(\theta_k,x_k)&\leq \oa(\|x_k\|) \\
		\Delta V(\theta_k,x_k)&\leq -\da(\|x_k\|). 
	\end{align}
\end{subequations}
Hence, \eqref{eq:fullLF} gives a Lyapunov function for $x_k$, i.e., according to LaSalle's invariance principle~\cite{hespanha2004uniform}, it guarantees AS of the state $x$, while the parameter $\theta$ evolves freely. 
\end{IEEEproof}

For the results of Theorem~\ref{th:AS} to hold, the parameter vector $\theta$ must be 
updated only when 
\eqref{eq:safetyCheck} is satisfied, which is an easy condition to check. The additional 
assumptions of~\eqref{eq:Kbounds} may be checked offline, for all parameter values 
$\theta\in\Theta$.  Alternatively, one could start from pre-assigned $\ua,\oa,\da\in\KI$, 
and accept a parameter update value $\ovl \theta$ only if~\eqref{eq:Kbounds} is satisfied 
for $\Theta\supseteq \{\theta_j\}^{k-1}_{j=0}\cup \ovl \theta$. 
Theorem~\ref{th:AS} does not provide guarantees for the convergence of $\theta$ to a value, and in fact the parameter often exhibits at steady state oscillation with a small amplitude, especially in the presence of disturbances or modeling errors. 
In practice, the calibration algorithm will freeze the parameter vector, i.e., it will stop the updates, when the updates or their impact is smaller than a used-defined threshold, to possibly restart later if the proposed updates grow again, e.g., due to some condition changes.

Finally, the condition~\eqref{eq:safetyCheck} in Theorem~\ref{th:AS} is only one example of the conditions that can be enforced by the safety check to ensure AS. 
We derived condition~\eqref{eq:safetyCheck} from switched Lyapunov functions, such as those for the analysis of piecewise affine systems~\cite{mignone2000stability}.  
Alternative, more relaxed conditions can be  applied from other methods in switched dynamical systems. 
For instance, when there exist Lyapunov functions for constant control parameters~\eqref{eq:constantLF} for all $\theta\in\Theta$,  dwell time methods may be applied to hold the parameter constant ``long enough'' to ensure that the Lyapunov function with changing parameter values decreases with respect to previous most recent parameter update. 
Thus, the value  $\theta_{\bar k}\neq \theta_{\bar k-1}$ is accepted only if 
$ V_{\theta_{\bar  k}}(x_{\bar k})\leq  V_{\theta_{\hat  k}}(x_{\hat  k}) 
-\alpha^\Delta(\|x_{\hat k}\|)$ 
where $\hat k= \max_{h\leq k-1}\{\theta_h:\ \theta_h\neq \theta_{h-1}\}$. 
This condition ensures stability according to the results in~\cite{liberzon1999stability}, through the stability of the subsequences at the instances of parameter updates.

\section{Simulation Results}
\label{sec:doubleInt}
\subsection{Application to Controlling Double Integrator}
In this section, we show the proposed Kalman filter-based calibration method for different controller architectures. 
For this purpose, we study a simple double-integrator system model, 
\begin{subequations}
\begin{align} 
\label{eq:doubleInt}
    \begin{bmatrix}
    p_{k+1}\\
    v_{k+1}
    \end{bmatrix}
    & =
    \begin{bmatrix}
    1 & T_s \\ 0 & 1
    \end{bmatrix}
    \begin{bmatrix}
    p_{k}\\
    v_{k}
    \end{bmatrix}
    +
    T_s
    \left(
    \begin{bmatrix}
    0\\
    1
    \end{bmatrix}
    u_k
    +
    \begin{bmatrix}
    0\\
    1
    \end{bmatrix}
    \delta v_{k}
    \right),
    \\
    e_k &= p_k - p_{\rm ref},
\end{align}
\end{subequations}
with the goal to track a certain reference position, $p_{\rm ref}$, where $p_k$ is the position at time step $k$, $v_k$ is the velocity, $u_k$ is the input (the acceleration), and $\delta v_{k}$ is unmodeled noise/disturbance.
The sampling period is $T_s=0.1$~s.
Throughout this paper, we choose ${\rm C}_\theta=I$ implying that there are no preferences for tuning specific parameters faster than others. Similarly, ${\rm C}_v= I$.

\subsubsection{Design Choices for Controller Architectures}
First, we show the parametric structure of the different controllers applied to the system in \eqref{eq:doubleInt}. For the calibration, we utilize the UKF implementation in Section~\ref{ssec:UKF}. 
\begin{itemize}
    \item 
State Feedback Control: For this controller, the Kalman filter calibrates the feedback gain directly, 
\begin{align*}
    u_k = 
    \begin{bmatrix}
        \theta_{1} & \theta_{2}
    \end{bmatrix}
    \begin{bmatrix}
        e_k\\ v_k
    \end{bmatrix},
\end{align*}
i.e., there are two control parameters.
\item 
Optimal Control:
For this controller, the Kalman filter calibrates the quadratic cost function,
\begin{align*}
    J = \sum_{k=0}^\infty 
    \begin{bmatrix}
        e_k\\ v_k
    \end{bmatrix}^T
    \begin{bmatrix}
    \theta_{1} & \theta_{2}
    \\
    \theta_{2} & \theta_{3}
    \end{bmatrix} 
    \begin{bmatrix}
        e_k\\ v_k
    \end{bmatrix}
    +
    \theta_4 u_k^2,
\end{align*}
which we implement as a linear–quadratic regulator, 
\begin{align*}
    u_k = 
    k_\theta 
    \begin{bmatrix}
        e_k\\ v_k
    \end{bmatrix},
\end{align*} 
where $k_\theta$ results from the discrete-time algebraic Riccati equation (DARE).
Hence, there are four control parameters.
\item 
PID Control: For this controller, the Kalman filter calibrates three control parameters, the proportional, the integral, and the derivative gain,
\begin{align*} 
    u_k =\ &
    u_{k-1} + 
    (\theta_P+\theta_I+\theta_D)e_k
    \\
    &
    + (\theta_I-\theta_P-2\theta_D)e_{k-1} + \theta_De_{k-2}.
\end{align*}  
\item
$H_\infty$ Control: As this controller is often implemented for output feedback, we assume that only the position is measurable. Here, we calibrate filter coefficients of a pre-compensator and a post-compensator of the dynamical system. 
This loop-shaping technique can be seen, e.g., in Fig.~9.17 in \cite{skogestad2005multivariable}, and is applied to obtain a trade-off between robustness (provided by the $H_\infty$ optimization) and performance. For the $H_\infty$ optimization we use the algorithm in Table~9.2 in \cite{skogestad2005multivariable}. 
We choose both pre- and the post-compensator as first-order filters
\begin{align*}
    H_{\rm pre}(s) = \frac{\theta_1s+\theta_2}{\theta_3s+\theta_4},\ 
    H_{\rm post}(s) = \frac{\theta_5s+\theta_6}{\theta_7s+\theta_8},
\end{align*}
i.e., there are eight control parameters. Then, the controller is given by
\begin{align*}
     z_{k+1} &= A_\infty  z_k + b_\infty e_k,
    \\
    u_k &= c_\infty  z_k + d_\infty e_k,
\end{align*}
where $A_\infty\in\mathbf{R}^{6\times 6},b_\infty\in\mathbf{R}^{6},c_\infty\in\mathbf{R}^{1\times 6},d_\infty\in\mathbf{R}$ result from the $H_\infty$ optimization algorithm in combination with the pre- and post-compensators and $z_k$ is the state of the controller.
\item 
Sliding Mode Control:
For the double integrator model, the control law becomes 
$$u_k = -\theta_1 v_{k} - \theta_2\cdot {\rm sign}(e_{k}+\theta_1 v_{k}),$$ 
where $e_{k}+\theta_1 v_{k}$ is the sliding surface,
i.e., there are two control parameters.  
\item 
Dynamic Output Feedback Control: For this controller, we assume that only the position is measurable and the controller is given by  
\begin{align*} 
    \begin{bmatrix}
        \hat p_{k+1} \\ \hat v_{k+1}
    \end{bmatrix} &= 
    \begin{bmatrix}
    1 & T_s \\ 0 & 1
    \end{bmatrix}
    \begin{bmatrix}
        \hat p_{k} \\ \hat v_{k}
    \end{bmatrix}
    + 
    \begin{bmatrix}
        0 \\ T_s
    \end{bmatrix} u_k + 
    \begin{bmatrix}
        \theta_3 \\ \theta_4
    \end{bmatrix} (\hat p_k - p_k),
    \\
    u_k &= 
    \begin{bmatrix}
        \theta_1 & \theta_2
    \end{bmatrix}
    \left( 
    \begin{bmatrix}
        \hat p_k - p_{\rm ref} \\ \hat v_k
    \end{bmatrix}
    \right),
\end{align*}
where $\hat p_k$ and $\hat v_k$ are the estimated position and velocity, respectively. 
Hence, there are four control parameters, where $\theta_1$, $\theta_2$ define the feedback gain and $\theta_3$, $\theta_4$ define the observer gain.  

\item 
Neural Network Control: 
We choose a fully connected Neural Network with one input layer ($2\cdot10\!+\!10\!=\!30$ parameters), one hidden layer ($10\cdot10\!+\!10\!=\!110$ parameters), and one output layer ($1\cdot10\!+\!1\!=\!11$ parameters), i.e., the control law is 
$$u_k=\theta_{\rm out}\sigma\left(\theta_{\rm lay}\sigma\left(\theta_{\rm in}
\begin{bmatrix}
    e_k \\ v_k
\end{bmatrix}
+\theta_{{\rm in},0}\right)+\theta_{\rm lay,0}\right)+\theta_{{\rm out},0}$$ 
with 
$\theta_{\rm out}\in \mathbf{R}^{1\times 10}$, $\theta_{{\rm out},0}\in \mathbf{R}$,
$\theta_{\rm lay}\in \mathbf{R}^{10\times 10}$,
$\theta_{\rm lay,0}\in \mathbf{R}^{10\times 1}$, 
$\theta_{\rm in}\in \mathbf{R}^{10\times 2}$, and
$\theta_{{\rm in},0}\in \mathbf{R}^{10\times 1}$.
We use the leaky ReLu activation function $y=\sigma(x)$ with $y_i=\max(0.1x_i,x_i)$ for all elements $i=1,...,10$.
Hence, the Kalman filter calibrates 151 parameters. 
\end{itemize} 

\subsubsection{Reference Position Tracking} 
\label{ssec:initialCal}
First, we study a reference tracking task, where $p_0=0$, $p_{\rm ref}=1$, and $\delta v_k=0$ with a time horizon of 15~s. 
In this simulation study, the proposed controller calibration method is implemented episodically, i.e., the Kalman filter is executed in between iterations/repetitions of the reference tracking task and 15~s of data. 
We use two training objectives,
\begin{subequations}
\begin{align}
\label{eq:training1}
    y_k &=
    \begin{bmatrix}
        p_{\rm ref}\cdot \mathbf{1}_{150\times 1}
        \\
        \mathbf{0}_{150\times 1}
    \end{bmatrix},\
    h(\theta_k)=
    \begin{bmatrix}
        p_{1:150}
        \\
        u_{1:150}
    \end{bmatrix},
    \\
\label{eq:training2}
    y_k &=
    \begin{bmatrix}
        p_{\rm ref}\cdot \mathbf{1}_{150\times 1}
        \\
        \mathbf{0}_{150\times 1}
        \\
        0
    \end{bmatrix},\
    h(\theta_k)=
    \begin{bmatrix}
        p_{1:150}
        \\
        u_{1:150}
        \\
        {\rm cost}_{\rm overshoot}
    \end{bmatrix},
\end{align}
\end{subequations}
with $p_{1:150}=[p_1\ p_2\ ...\ p_{150}]^T,\ u_{1:150}=[u_1\ u_2\ ...\ u_{150}]^T\in \mathbf{R}^{150}$, and 
\begin{align*}
    {\rm cost}_{\rm overshoot}=\begin{cases}
    10 & {\rm if}\ \max(p_{1:150})>1.1
    \\
    0 & {\rm else}.
    \end{cases}
\end{align*}
Note that \eqref{eq:training1} and \eqref{eq:training2} differ only by virtue of the additional cost for overshooting.

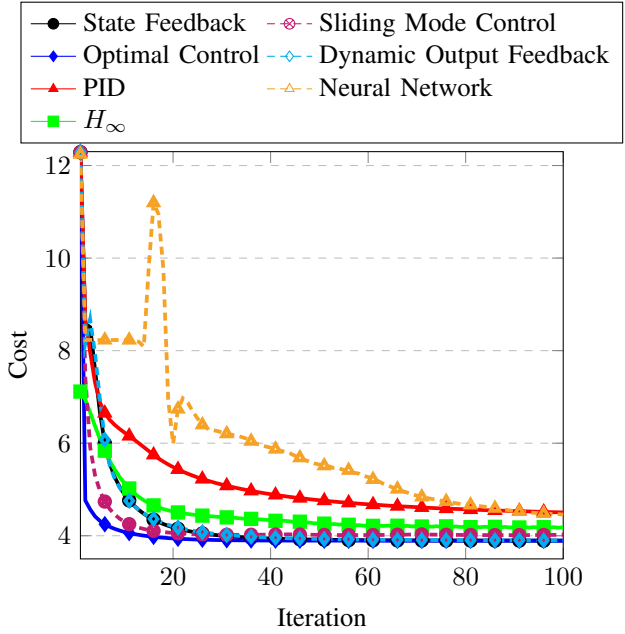
\begin{figure}
    \centering 
\begin{tikzpicture}[every plot/.append style={line width=0.5mm},every mark/.append style={solid},]
\begin{axis}[
    xmin=1, xmax=100,
    ymin=3.5, ymax=12.3,  
    %ymode=log,
   ylabel={Cost}, 
   xlabel={Iteration}, 
  % yticks = {5,10}, 
   %yticklabels={ },  
   %xticklabels={,,,,,,,,,,,,,,,}, 
    legend style={at={(0.5,1.02)},anchor=south},  
    ymajorgrids=true,
    legend columns=4,
    transpose legend,
    grid style=dashed,
    width=8cm, 
    height=7.0cm,
    mark repeat=5,
    y label style={at={(axis description cs:0.1,.5)},rotate=0,anchor=south},
    legend cell align={left}
	] 
]	  
\addplot[color=black,mark=*] table[x index=0,y index=1] {data/journalConv.txt};
\addplot[color=blue,mark=diamond*] table[x index=0,y index=2] {data/journalConv.txt};
\addplot[color=red,mark=triangle*] table[x index=0,y index=3] {data/journalConv.txt};
\addplot[color=green,mark=square*] table[x index=0,y index=4] {data/journalConv.txt};
\addplot[color=mst3,densely dashed,mark=otimes] table[x index=0,y index=5] {data/journalConv.txt};
\addplot[color=cyan,densely dashed,mark=diamond] table[x index=0,y index=6] {data/journalConv.txt};
\addplot[color=mst2,densely dashed, mark=triangle] table[x index=0,y index=7] {data/journalConv.txt};

\legend{State Feedback,Optimal Control,PID,$H_\infty$,Sliding Mode Control,Dynamic Output Feedback,Neural Network};

\end{axis}
\end{tikzpicture} 
    \caption{Convergence of closed-loop cost for seven controllers. The plot displays the initial calibration phase of the controllers, where the Kalman filter is implemented episodically, i.e., the reference tracking task is executed 100 times. The model-based calibration enables fast convergence for all controller structures.
    }
    \label{fig:convergenceStudy}
\end{figure}
 
Fig.~\ref{fig:convergenceStudy} shows the convergence of the closed-loop cost for the seven different controllers using the training objective~\eqref{eq:training1}.
It shows that all controllers gradually decrease the cost, but the convergence speed and achieved cost after 100 iterations differ.
The optimal controller achieves both the fastest convergence and the lowest steady-state cost, which is expected due to the structural similarities between the cost function and the training objective.
The neural network controller converges the slowest, which is also expected due to the high number of parameters and its black-box design.
The $H_\infty$ controller converges to a slightly higher cost than some of the other controllers, because the learning method balances performance with robustness. 
The $H_\infty$ controller also requires an initial set of parameters that can stabilize the system, which is the reason why the initial cost is lower than for the other controllers.
The PID controller also converges to a slightly higher cost, which is caused by the integral term, which requires the controller to overshoot the reference value.
In order to quantify convergence speed, we empirically compute decay factors for the seven controllers as 
\begin{align}
\label{eq:decay}
\Delta F 
=
    \frac{1}{99}\sum_{i = 0}^{98}
\frac{\bar c_{i}-\bar c_{i+1}}{\bar c_{i+1}},\quad
    \bar{c}_i = \frac{c_i-c_{100}}{c_0-c_{100}},
\end{align}
where $c_i$ is the cost at iteration $i$ as in Fig.~\ref{fig:convergenceStudy}.
The decay factor, $\Delta F$, evaluates the average of how much the cost is reduced at each iteration. In Fig.~\ref{fig:convergenceStudy}, the decay factors as in \eqref{eq:decay} are 16.4\% (state feedback), 79.2\% (optimal control), 6.55\% (PID), 7.82\% ($H_\infty$), 37.6\% (sliding mode control), 15.3\% (dynamic output feedback), and 6.39\% (neural network). On average, the decay factor is 24.2\%.
 
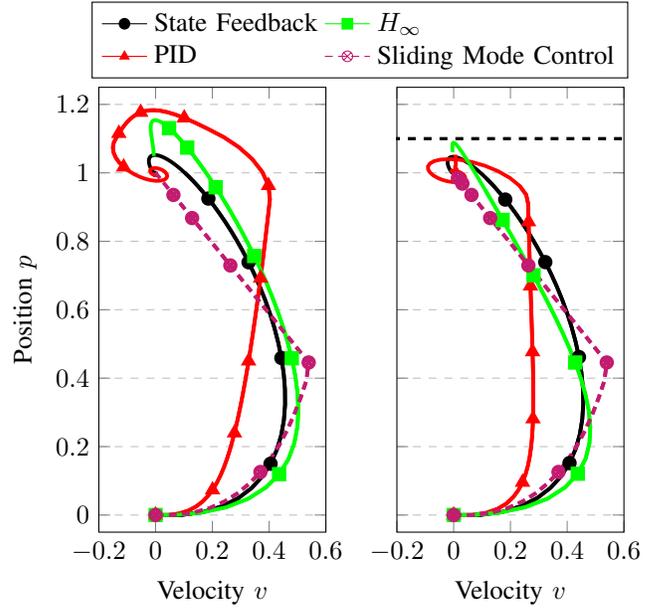
\begin{figure}
    \centering 
\begin{tabular}{rr} \begin{tikzpicture}[every plot/.append style={line width=0.5mm},every mark/.append style={solid},]
\begin{axis}[
    xmin=-0.2, xmax=.6,
    ymin=-0.05, ymax=1.25,   
   xlabel={Velocity $v$}, 
   ylabel={Position $p$},  
    legend style={at={(1.15,1.02)},anchor=south},  
    ymajorgrids=true,
    legend columns=2,
    transpose legend,
    grid style=dashed,
    width=4.6cm, 
    height=7.5cm,
    mark repeat=7,
    y label style={at={(axis description cs:0.175,.5)},rotate=0,anchor=south},
    legend cell align={left}
	] 
]	  
\addplot[color=black,mark=*] table[x index=1,y index=0] {data/StateSpace_FSF_noOver_MARKS.txt};  
\addplot[color=red,mark=triangle*] table[x index=1,y index=0] {data/StateSpace_PID_noOver_MARKS.txt};

\addplot[color=green,mark=square*] table[x index=1,y index=0] {data/StateSpace_Hinf_noOver_MARKS.txt}; 

\addplot[color=mst3,densely dashed,mark=otimes] table[x index=1,y index=0] {data/StateSpace_SMC_noOver_MARKS.txt};  

\addplot[color=black] table[x index=1,y index=0] {data/StateSpace_FSF_noOver.txt}; 
\addplot[color=red] table[x index=1,y index=0] {data/StateSpace_PID_noOver.txt};
\addplot[color=green] table[x index=1,y index=0] {data/StateSpace_Hinf_noOver.txt}; 
\addplot[color=mst3,densely dashed] table[x index=1,y index=0] {data/StateSpace_SMC_noOver.txt};

\legend{State Feedback,PID,$H_\infty$,Sliding Mode Control}; 
     
\end{axis}
\end{tikzpicture}    
\hspace{-3.8cm}
\begin{tikzpicture}[every plot/.append style={line width=0.5mm},every mark/.append style={solid},]
\begin{axis}[
    xmin=-0.2, xmax=.6,
    ymin=-0.05, ymax=1.25,   
   xlabel={Velocity $v$},  
   yticklabels={ },  
    legend style={at={(0.5,1.02)},anchor=south},  
    ymajorgrids=true,
    legend columns=4,
    transpose legend,
    grid style=dashed,
    width=4.6cm, 
    height=7.5cm,
    mark repeat=7,
    y label style={at={(axis description cs:0.05,.5)},rotate=0,anchor=south},
    legend cell align={left}
	] 
]	  
\addplot[color=black,mark=*] table[x index=1,y index=0] {data/StateSpace_FSF_Over_MARKS.txt}; 
\addplot[color=black] table[x index=1,y index=0] {data/StateSpace_FSF_Over.txt};  
\addplot[color=red,mark=triangle* ] table[x index=1,y index=0] {data/StateSpace_PID_Over_MARKS.txt};
\addplot[color=red] table[x index=1,y index=0] {data/StateSpace_PID_Over.txt};

\addplot[color=green,mark=square* ] table[x index=1,y index=0] {data/StateSpace_Hinf_Over_MARKS.txt}; 
\addplot[color=green] table[x index=1,y index=0] {data/StateSpace_Hinf_Over.txt}; 

\addplot[color=mst3,densely dashed,mark=otimes ] table[x index=1,y index=0] {data/StateSpace_SMC_Over_MARKS.txt}; 
\addplot[color=mst3,densely dashed] table[x index=1,y index=0] {data/StateSpace_SMC_Over.txt};

\draw[dashed,very thick] (axis cs:-1,1.1) -- (axis cs:1.0,1.1);

\end{axis}
\end{tikzpicture}   
\end{tabular}
    \caption{
    State trajectories for selection of controllers.
    Left: State trajectories after calibration using training objective \eqref{eq:training1}. 
    Right: State trajectories after calibration using training objective \eqref{eq:training2} with cost for overshoot.
    The full state feedback controller and the sliding mode controller are not changed by the additional penalty for overshooting, since they do not overshoot to begin with. 
    The sliding surface can be identified easily from this plot. On the other hand, the state trajectories of the PID and the $H_\infty$ controller change to avoid overshooting.
    }
    \label{fig:stateSpace}
\end{figure}

Fig.~\ref{fig:stateSpace} illustrates a selection of state trajectories of different controllers with the training objective in \eqref{eq:training1} and \eqref{eq:training2}. 
The selection of controllers is made to illustrate the implications of the two training objectives.
While some controllers avoid overshooting with training objective \eqref{eq:training1}, even without the additional penalty, see state feedback and sliding mode control, some controllers overshoot the reference value by more than 10\%, see PID and $H_\infty$ control. 
However, if a penalty for overshooting is added as in \eqref{eq:training2}, then all controllers limit overshooting the reference position to a maximum value of 1.1. 
In particular, the state feedback and sliding mode controller's performances remain unchanged, while the performances of the PID and the $H_\infty$ controller are altered to meet the specification.

\subsubsection{Continuous Control Task with Disturbances}
\begin{table*}[t]
    \centering
    \caption{
    Cost for Different Process Noise and Disturbances
    }
    \begin{tabular}{ll|cc|cccccccccccccc}  
        \toprule
    & &\multicolumn{4}{c}{Test Case for Regulation} \\
         & &  \multicolumn{2}{c}{$\delta v_{k}\sim \mathcal{N}(0,1)$} & \multicolumn{2}{c}{$\delta v_{k}=1$} \\
        Controller Type & Parameters that are adjusted &  Initial Param. &  Param. adjusted online &  Initial Param. &  Param. adjusted online
        \\
        \midrule
        State Feedback & Feedback gain &
        0.166 & 0.166 (-0.01\%) & 0.470 & 0.413(-12.1\%)
        \\
        Optimal Control & Cost function weights &
        0.173 & 0.173 (-0.01\%) & 0.465 & 0.436 (-6.24\%)
        \\
        PID &  Gains &
        0.367 & 0.352 (-4.08\%) & 0.536 & 0.361 (-32.7\%)
        \\
        $H_\infty$ & Filter coefficients  &
        0.380 & 0.300 (-21.1\%) & 5.67 & 0.856 (-84.9\%) 
        \\
        Sliding Mode Control &  Sliding surface and gains  &
        0.489 & 0.425 (-13.2\%) & 0.318 & 0.318 (-0.01\%) 
        \\
        Dynamic Output Feedback & Feedback and observer gains  &
        2.95 & 2.94 (-0.34\%) & 3.04 & 2.82 (-7.15\%)
        \\
        Neural Network & Weights &
        6.32 & 2.33 (-63.2\%) & 6.01 & 2.33 (-61.9\%)  
        \\
        \bottomrule
    \end{tabular}
    \label{tab:noise}
\end{table*}
Table~\ref{tab:noise} shows a simulation study in which the controllers are exposed to noise/disturbances. 
It shows two test cases, $\delta v_{k}\sim \mathcal{N}(0,1)$ and $\delta v_{k}=1$, and the controllers aim at regulating the system around the origin, $p_{\rm ref}=p_0=0$. 
In this simulation study, the proposed controller calibration method is implemented online, i.e., the Kalman filter is executed in receding horizon fashion at each time step with a sliding window of length $N$.
The training objective for this task is~\eqref{eq:training1}.
The table shows a comparison between two sets of control parameters for each controller and each test case. The first set of control parameters is the result of the initial calibration obtained after 100 iterations as in Section~\ref{ssec:initialCal}. The second set of control parameters uses online calibration (starting from the same initial parameters) to improve upon the closed-loop system performance.
It can be seen that no controller performs worse using the online adaptation in any test case. 
This indicates that the Kalman filter does not over-fit the parameters in the presence of noise. 
While all controllers improve their performance, the $H_\infty$ controller exhibits one of the largest improvements. 
This makes sense as the pre- and post-compensators are adjusted to exhibit the desired frequency response by the Kalman filter.
For the neural network controller, the performance improves by more than 60\%, which is expected as the neural network adjusts to the specific test case and because more data are used for training.

\subsection{Application to Vehicle Model and Comparison of UKF, EKF, and Bayesian Optimization}
In the following, we present a comparison of UKF, EKF, and Bayesian optimization applied to a reference tracking task using an optimal controller, in which a vehicle tracks a certain lateral deviation from the center-line and a certain velocity.
Here, the proposed method is implemented episodically similar to BO.
The kinematic single track vehicle model is given by~\cite{Carvalho2015} 
\begin{subequations}
\label{eq:kinematicBici}
\begin{align}
    x_{k+1} = f(
    x_k,
    u_k)
\end{align}
with 
$
    x_k = \begin{bmatrix}
p_{X,k}
&
p_{Y,k}
&
\psi_k
&
v_k
&
\delta_k
\end{bmatrix}^T, u_k = \begin{bmatrix}
\dot v_k
&
\dot \delta_k
\end{bmatrix}^T
$, 
and 
\begin{align}
    f(x_k,u_k) = \begin{bmatrix}
    v_{x,k} \cos(\psi_k+\beta_k)/\cos(\beta_k)
    \\
    v_{x,k} \sin(\psi_k+\beta_k)/\cos(\beta_k)
    \\
    v_{x,k} \tan(\delta_k)/L
    \\
    \dot v_k
    \\
    \dot \delta_k
    \end{bmatrix},
\end{align}
where $p_X$ and $p_Y$ are the vehicle's position in the world frame, 
$\psi$ is the heading (yaw) angle,
$v_x$ is the longitudinal velocity,
$\delta$ is the steering angle of the front wheel,
$L=l_f+l_r$ is the wheel base, and
$\beta= \arctan(l_r \tan(\delta)/L)$ is the kinematic body-slip angle.
The inputs $u_1$ and $u_2$ are the longitudinal acceleration and the steering rate.
The road is oriented such that the longitudinal progress is in the $p_X$ direction. 
\end{subequations}

We use the quadratic cost function $\!\sum_{k=0}^{N\!-\!1}\|M(x_k-x_{\rm ref})\|_{Q}\!+\!\|u_k\|_R$ with $M\!=\![0_{4\times 1}\ I_4]$, i.e., all states but the longitudinal progress are penalized (road oriented in $p_X$ direction).
Hence, we learn the cost function parameters $Q\!\in\!\mathbf{R}^{4\times 4}$ and $R\!\in\! \mathbf{R}^{2\times 2}$. %, where $Q,R\!\succeq\! \epsilon \cdot I$, with a small $\epsilon$.
Throughout this section, we use the sampling time $T_s=0.25$~s and a planning horizon of the optimal controller $N=20$.
For all simulations and methods in the following, we initialized the parameters, $Q,R$, using randomly sampled positive definite matrices.

\subsubsection{Comparison with Bayesian Optimization}
\label{sec:comp_BO}
\begin{figure}[t]
    \centering 
    \begin{tabular}{cc}
    \begin{tikzpicture}[every plot/.append style={thick}]
\begin{axis}[
    xmin=0, xmax=59,
    ymin=-.1, ymax=2.2, 
   ylabel={Cost},  
   xticklabels={,,,,,,,,,,,,,,,}, 
    legend style={at={(1,1)},anchor=north east},  
    ymajorgrids=true,
    legend columns=1,
    transpose legend,
    grid style=dashed,
    width=8cm, 
    height=4cm,
    %legend cell align={left}
	] 
]	 
\addplot[gray,fill=gray!50,opacity=.7] table[x index=0,y index=1] {data/ekf_diag_std.txt};
\addplot[color=black] table[x index=0,y index=1] {data/ekf_diag_mean.txt}; 
\addplot[red,fill=red!50,opacity=.3] table[x index=0,y index=1] {data/ukf_diag_std.txt};
\addplot[color=red,densely dashed] table[x index=0,y index=1] {data/ukf_diag_mean.txt};
\addplot[gray,fill=green!50,opacity=.2] table[x index=0,y index=1] {data/bo_std.txt};
\addplot[densely dashdotted,color=green,] table[x index=0,y index=1] {data/bo_mean.txt}; 

\legend{,Extended Kalman Filter, ,Unscented Kalman Filter,,Bayesian Optimization};
    
\end{axis}
\end{tikzpicture}
\\
\begin{tikzpicture}[every plot/.append style={ thick}]
\begin{axis}[
    xmin=0, xmax=59,
    ymin=-.1, ymax=2.2, 
   ylabel={Average Cost}, 
   xlabel={Iteration}, 
    ymajorgrids=true,
    legend columns=1,
    transpose legend,
    grid style=dashed,
    width=8cm, 
    height=4cm,
	] 
]	 
\addplot[gray,fill=gray!50,opacity=.7] table[x index=0,y index=1] {data/ekf_diag_cum_ave_std.txt};
\addplot[color=black,] table[x index=0,y index=1] {data/ekf_diag_cum_ave_mean.txt}; 
\addplot[red,fill=red!50,opacity=.3] table[x index=0,y index=1] {data/ukf_diag_cum_ave_std.txt};
\addplot[color=red,densely dashed,] table[x index=0,y index=1] {data/ukf_diag_cum_ave_mean.txt}; 
\addplot[gray,fill=green!50,opacity=.2] table[x index=0,y index=1] {data/bo_cum_ave_std.txt};
\addplot[color=green,densely dashdotted,] table[x index=0,y index=1] {data/bo_cum_ave_mean.txt}; 
     
\end{axis}
\end{tikzpicture} 
\end{tabular}
    \caption{Evaluation of controller performance. Both recursive algorithms, EKF and UKF, as well as BO are displayed. The plot shows the median (solid or dashed lines), as well as 75th and 25th percentiles (shaded areas) of 500 trials. Top: Cost of operation at each time step.  
    Bottom: Average cost of controllers over time. The bottom plot is particularly interesting as it shows the increased cost that BO incurs due to the trial-and-error implementation. 
    }
    \label{fig:my_label}
\end{figure}
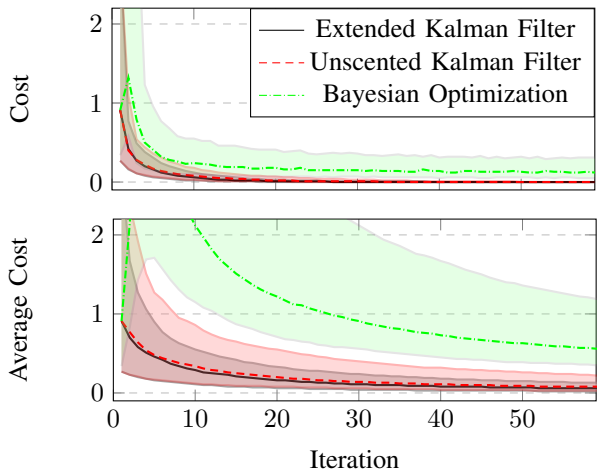

We consider a reference tracking task with $x_{\rm ref}\!=\![0,\,0,\,0,\,10~{\rm m/s},\,0]^T$, where the initial state is $x_0\!=\![
0,\,
2~{\rm m},\,
0,\,
12~{\rm m/s},\,
0
]^T$.
The learning algorithm uses the training objective
\begin{align*} 
y_k=0,\quad 
    h_1(\theta)= 
    \begin{bmatrix}
    M(x_0-x_{\rm ref}) \\ \vdots \\ M(x_N-x_{\rm ref})
    \\
    u_0 \\ \vdots \\ u_{N-1}
    \end{bmatrix}.
\end{align*}
Fig.~\ref{fig:my_label} reports the cost $\|y_k-h(\theta)\|_2^2$ over time, by showing the results of 500 trials, where the cost parameters are initialized randomly at the beginning of each trial.
It shows the convergence behaviors of the two recursive algorithms proposed in this paper, the EKF and the UKF, along with a Bayesian optimization (BO) approach, which is added as comparison.
Here we use an episodical learning task to obtain a meaningful comparison with BO.
The BO approach uses a squared exponential kernel and the Upper Confidence Bound (UCB) acquisition function, both commonly used in the literature, e.g.,~\cite{lu2020mpc}.
Further, we restrict the BO approach also to the search space of positive definite matrices.
It can be seen that both the EKF and UKF variants outperform BO in terms of convergence speed and steady-state cost at the end of the simulation at time step 100.  
We used the same randomly sampled positive definite matrices for the BO as for the Kalman filter-based adaptation. For BO, we used the squared exponential kernel with lengthscale $l\!=\!1$ and output variance $\sigma^2\!=\!1$.

\begin{remark}
The slower convergence speed of BO can be attributed to its trial-and-error approach, in which BO keeps testing control parameters that have not been tested in order to ensure global optimality. 
Other BO implementations in the literature may have an improved convergence speed.
For example, there are BO approaches that use gradient information~\cite{wu2017bayesian} or focus on constraining the policy search space to a local domain~\cite{frohlich2021cautious}. 
For some applications, a potential drawback of the method proposed in this paper can be that the lack of global exploration, compared to BO. 
Combining the method in this paper with BO to achieve both fast and global convergence may be subject to future studies.
\end{remark}

\subsubsection{EKF and UKF for Non-Differentiable Objective}
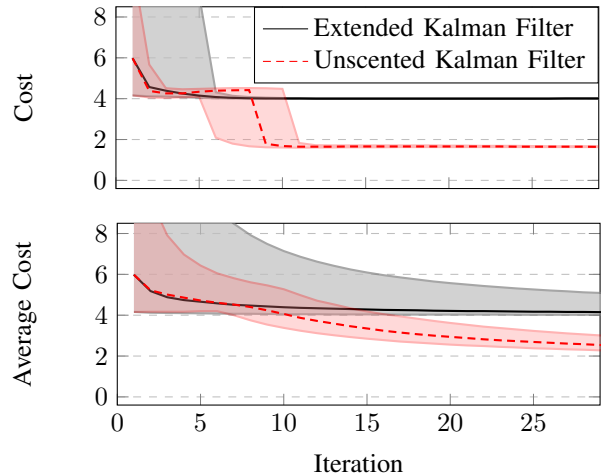
\begin{figure}[t]
    \centering  
\begin{tabular}{cc}  
\begin{tikzpicture}[every plot/.append style={ thick}]
\begin{axis}[
    xmin=0, xmax=29,
    ymin=-.4, ymax=8.5, 
   ylabel={Cost}, 
   xticklabels={,,,,,,,,,,,,,,,}, 
    legend style={at={(1,1)},anchor=north east}, 
    ymajorgrids=true,
    legend columns=1,
    transpose legend,
    grid style=dashed,
    width=8cm, 
    height=4cm,
    legend cell align={left}
	] 
]	 
\addplot[gray,fill=gray!50,opacity=.7] table[x index=0,y index=1] {data/ekf_nondiff_std.txt};
\addplot[color=black,] table[x index=0,y index=1] {data/ekf_nondiff_mean.txt}; 
\addplot[red,fill=red!50,opacity=.3] table[x index=0,y index=1] {data/ukf_nondiff_std.txt};
\addplot[color=red,densely dashed,] table[x index=0,y index=1] {data/ukf_nondiff_mean.txt};
\legend{,Extended Kalman Filter, ,Unscented Kalman Filter};
    
\end{axis}
\end{tikzpicture} 
\\
\begin{tikzpicture}[every plot/.append style={ thick}]
\begin{axis}[
    xmin=0, xmax=29,
    ymin=-.4, ymax=8.5, 
   ylabel={Average Cost}, 
   xlabel={Iteration}, 
    ymajorgrids=true,
    legend columns=1,
    transpose legend,
    grid style=dashed,
    width=8cm, 
    height=4cm,
	] 
]	 
\addplot[gray,fill=gray!50,opacity=.7] table[x index=0,y index=1] {data/ekf_cum_ave_nondiff_std.txt};
\addplot[color=black,] table[x index=0,y index=1] {data/ekf_cum_ave_nondiff_mean.txt}; 
\addplot[red,fill=red!50,opacity=.3] table[x index=0,y index=1] {data/ukf2_cum_ave_nondiff_std.txt};
\addplot[color=red,densely dashed] table[x index=0,y index=1] {data/ukf2_cum_ave_nondiff_mean.txt}; 
    
\end{axis}
\end{tikzpicture} 
\end{tabular}
    \caption{Evaluation of controller performance for partially non-differentiable objective.
    The plot shows the median (solid or dashed lines), as well as 75th and 25th percentiles (shaded areas) of 500 trials. 
    Top: Cost of operation at each time step.  
    Bottom: Average cost of controllers over time.
    The UKF implementation is able to identify the cost-beneficial tuning of the parameters. It can be seen that the UKF moves toward the beneficial region, even if it means to encounter slightly higher cost temporarily (top plot, around time steps 4--8).}
    \label{fig:nondiff}
\end{figure}

In addition to the tracking task described in Section~\ref{sec:comp_BO}, the training objective in this simulation includes a penalty for each sign change in~$\dot \delta$ with $y_k=0$ and $h_2(\theta)=[h_1(\theta)^T,\, \#\dot \delta\,{\rm sign\ changes}]^T$. 
One rational behind this penalty is to reduce oscillatory or jittering motions of the vehicle.
This additional penalty is a discontinuous function of the control input sequence.
Fig.~\ref{fig:nondiff} shows the performance of the EKF and UKF implementation in the presence of the partially non-differentiable training objective.
It shows that the UKF adjusts the parameters to capture the discontinuity, whereas the EKF converges to a set of parameters that incur higher cost.

\section{Application to Vehicle Steering Control in a High-Precision Simulation}
\label{sec:carsim}
This section presents the proposed calibration scheme  applied to calibrate two vehicle lane-change controllers implemented in CarSim~\cite{carsim}. 
CarSim is a high-fidelity vehicle dynamics simulator, which we utilize to study the method's performance and applicability to a complex dynamical system---with a simplified model for controller design. 
The rationale for this study is that a controller applied to a real-world system has to deal with a mismatch between the controller model and the physical system. 
CarSim is a third-party multi-body high-fidelity simulator for automotive applications, especially vehicle dynamics, that includes models for chassis dynamics, suspension, tires, road friction and slip, etc. 
On the other hand, the controller and the calibration module use a simple kinematic single track vehicle model. 
Hence, being able to execute the applied algorithms using a simplified control-oriented model against the high-fidelity simulator indicates the robustness of the method and suggests the applicability to a physical vehicle.

\subsection{CarSim Simulation Setup}
\begin{figure}
    \centering  
\includegraphics[width=.925\columnwidth]{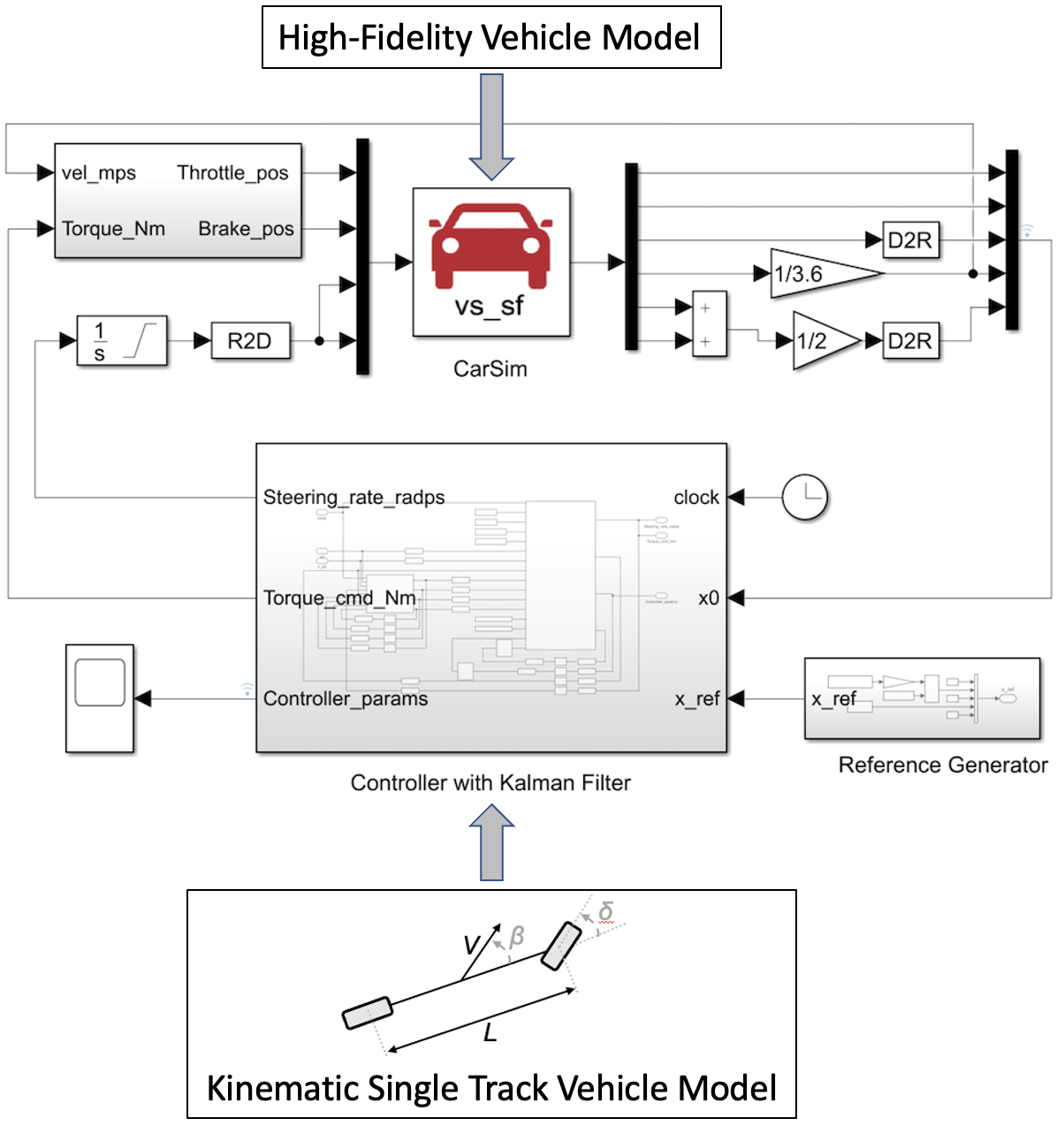}
    \caption{CarSim simulation setup with highlight on vehicle system model. Inputs to CarSim are the throttle position, the brake pedal position, as well as the steering angle command for the right and the left front wheels. Outputs from CarSim are the longitudinal position, the lateral position, the yaw angle, the vehicle velocity, as well as the steering angle for left and right wheels. The controller block uses the state (for a kinematic bicycle model) and a state reference in order to compute a steering rate command as well as a torque command. The steering angle is obtained by integrating the steering rate and the throttle and brake positions are computed using the torque command as well as the current velocity.
    }
    \label{fig:carsimInterface}
\end{figure}

\begin{figure}
    \centering 
\includegraphics[width=0.95\columnwidth]{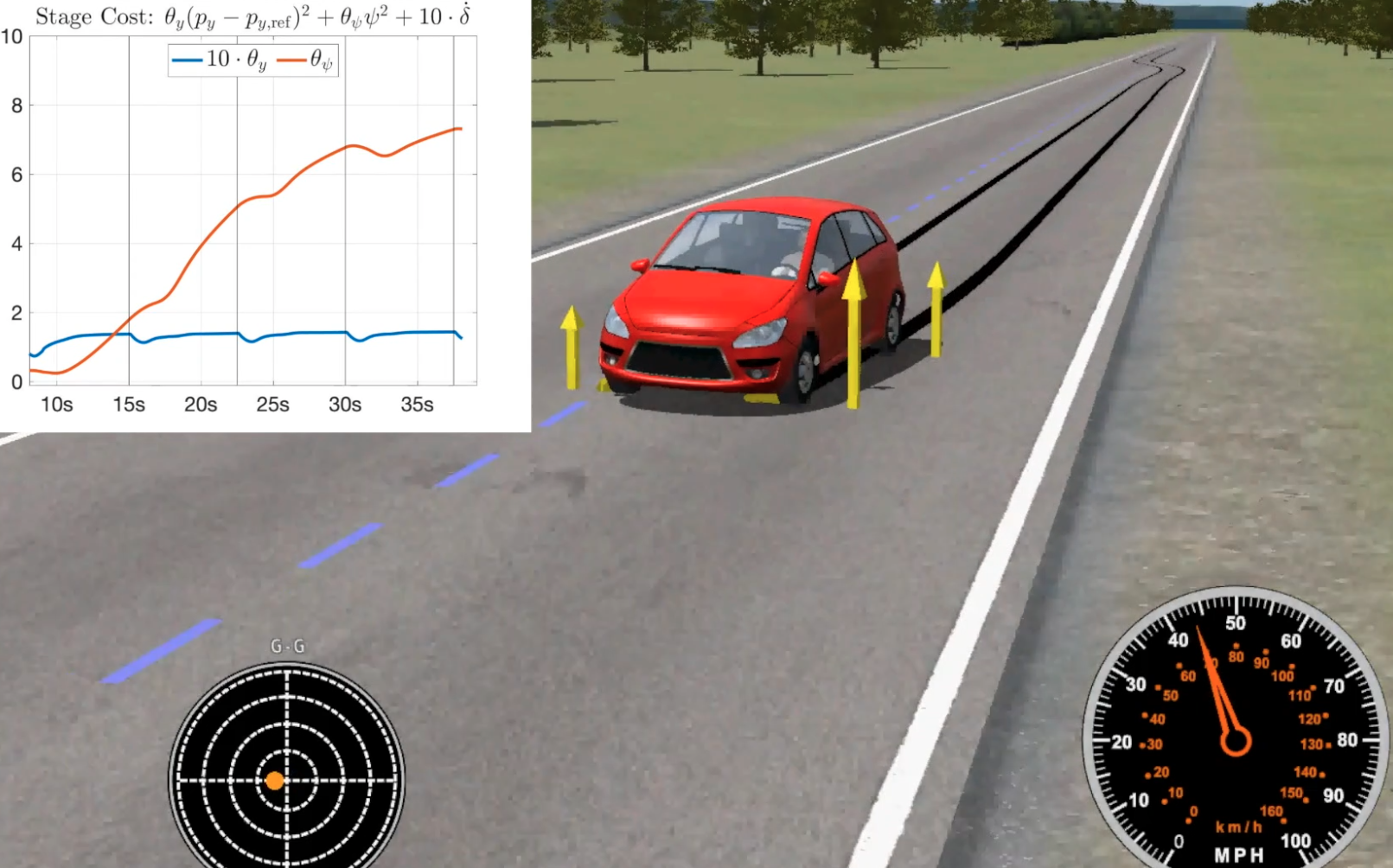}
    \caption{Illustration of vehicle performing lane-change maneuvers in CarSim, while controller parameters are adjusted online. 
    The yellow arrows illustrate forces acting onto the wheels. 
    }
    \label{fig:carsimVideo}
\end{figure}

Fig.~\ref{fig:carsimInterface} shows a block-diagram of a vehicle in CarSim being controlled in a Simulink interface. 
Fig.~\ref{fig:carsimVideo} shows a screenshot of the CarSim environment, in which a vehicle performs lane changes while adjusting its control parameters.
We have included a supplementary video to illustrate this scenario.
For reproducibility, the selected modules and simulation parameters can be found in the Appendix. 
The simulation sampling period 0.5~ms was chosen to be much faster than the control loop sampling period of 0.1~s, to make the simulation more precise.
The reference generator uses a pulsating signal, which switches between the lateral position references $\pm 1.5$~m every 7.5~s. 
We use a simple velocity controller to regulate the velocity as this study focuses on adjusting the lane-change controller.

\subsection{Controller and Calibration}
We implemented two types of controllers, an optimal controller and a state feedback controller. In the former, we use a kinematic single track vehicle model for both optimal control and the calibration of the control parameters. In the latter, a system model is not needed for the controller, but for the calibration of the control parameters. 
Fig.~\ref{fig:carsimInterface} also highlights the different vehicle system models used in the simulation setup.
While the optimal control and the controller calibration module use a kinematic single track vehicle model, the dynamical system to be controlled uses a high-fidelity vehicle model in CarSim.

We utilized the UKF implementation in Section~\ref{ssec:UKF}.
The UKF is implemented online in receding horizon fashion with a sliding window of length $N$.
The kinematic single track vehicle model is given as in \eqref{eq:kinematicBici}.
We use the training objective
\begin{align}
\label{eq:trainingCARSIM}
    y_k &=
    \begin{bmatrix}
        p_{Y,\rm ref}\cdot \mathbf{1}_{50\times 1}
        \\
        \mathbf{0}_{50\times 1}
        \\
        \mathbf{0}_{50\times 1}
    \end{bmatrix},\
    h(\theta_k)=
    \begin{bmatrix}
        p_{Y,1:50}
        \\
        0.1\cdot \psi_{1:50}
        \\
        10 \cdot u_{1:50}
    \end{bmatrix}.
\end{align}

\begin{remark}
It is possible to use a more complex vehicle model for the calibration and/or the controller, e.g., that includes the lateral dynamics of the vehicle based on tire forces. 
However, the presented results highlight the capabilities of the Kalman filter in the presence of a significant model mismatch, which will also be vital for real-world applications. 
\end{remark}

\subsection{Results for Optimal Controller}
\label{ssec:oc_results}
First, we analyze an optimal lane-change controller with the cost function
\begin{align}
\label{eq:lane_oc}
    J
    =
    \sum_{k=0}^{\infty}
    \theta_y p_{Y,k}^2
    +
    \theta_\psi \psi_{k}^2 
    +
    \dot \delta_k^2.
\end{align}
For simplicity, we linearize the model of the system dynamics in \eqref{eq:kinematicBici} around $x_{\rm lin}=[0\ 0\ 0\ v_x\ 0]^T$ and $u_{\rm lin}=[0\ 0]^T$ for computing the control law, which is obtained from solving the DARE. 
The UKF  uses the nonlinear system model in \eqref{eq:kinematicBici}. 

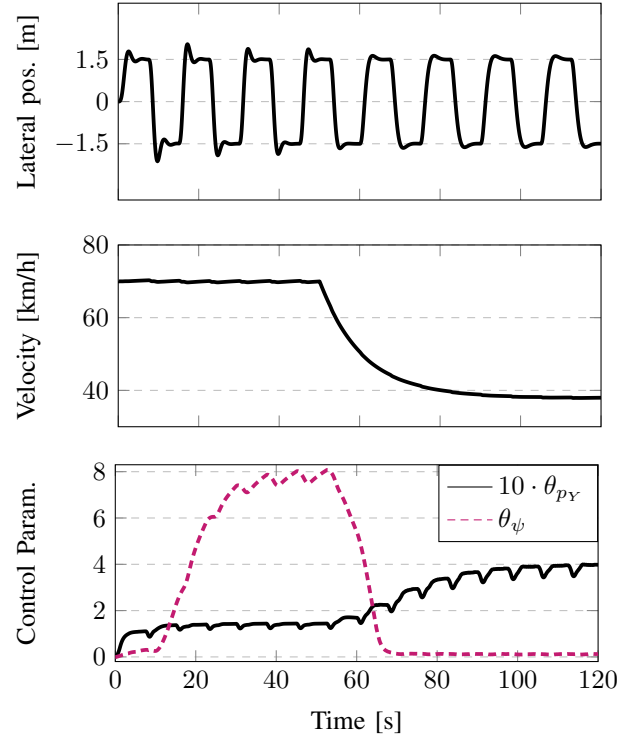
\begin{figure}
    \centering 
\begin{tabular}{ll} 
\begin{tikzpicture}[every plot/.append style={line width=0.5mm},every mark/.append style={solid},]
\begin{axis}[
    xmin=0, xmax=120,
    ymin=-3.5, ymax=3.5,   
   ylabel={Lateral pos. [m]},  
   ytick = {-1.5,0,1.5}, 
   xticklabels={,,,,,,,,,,,,,,,}, 
    legend style={at={(0.5,1.02)},anchor=south},  
    ymajorgrids=true,
    legend columns=4,
    transpose legend,
    grid style=dashed,
    width=8cm, 
    height=4.2cm,
    mark repeat=5,
    y label style={at={(axis description cs:0.05,.5)},rotate=0,anchor=south},
    legend cell align={left}
	] 
]	  
\addplot[color=black] table[x index=0,y index=1] {data_CarSim/lqr_pY.txt};   
  
\end{axis}
\end{tikzpicture} 
\\
\begin{tikzpicture}[every plot/.append style={line width=0.5mm},every mark/.append style={solid},]
\begin{axis}[
    xmin=0, xmax=120,
    ymin=30, ymax=80,   
   ylabel={Velocity [km/h]},  
  % yticks = {5,10}, 
   %yticklabels={ },  
   xticklabels={,,,,,,,,,,,,,,,}, 
    legend style={at={(0.5,1.02)},anchor=south},  
    ymajorgrids=true,
    legend columns=4,
    transpose legend,
    grid style=dashed,
    width=8cm, 
    height=4cm,
    mark repeat=5,
    y label style={at={(axis description cs:0.05,.5)},rotate=0,anchor=south},
    legend cell align={left}
	] 
]	     
\addplot[color=black] table[x index=0,y index=1] {data_CarSim/lqr_vel.txt}; 
  
\end{axis}
\end{tikzpicture} 
\\
\begin{tikzpicture}[every plot/.append style={line width=0.5mm},every mark/.append style={solid},]
\begin{axis}[
    xmin=0, xmax=120,
    ymin=-.2, ymax=8.3,   
   ylabel={Control Param.}, 
   xlabel={Time [s]}, 
  % yticks = {5,10}, 
   %yticklabels={ },  
   %xticklabels={,,,,,,,,,,,,,,,}, 
    legend style={at={(1,1)},anchor=north east},  
    ymajorgrids=true,
    legend columns=4,
    transpose legend,
    grid style=dashed,
    width=8cm, 
    height=4.2cm,
    mark repeat=5,
    y label style={at={(axis description cs:0.045,.5)},rotate=0,anchor=south},
    legend cell align={left}
	] 
]	     
\addplot[color=black] table[x index=0,y index=1] {data_CarSim/lqr_theta.txt}; 
\addplot[densely dashed,color=mst3] table[x index=0,y index=2] {data_CarSim/lqr_theta.txt};

\legend{$10\cdot \theta_{p_Y}$,$\theta_{\psi}$};
 
\end{axis}
\end{tikzpicture} 
\end{tabular}
    \caption{Calibration of optimal controller. A velocity-dependence of the control parameters can be seen as they change when the velocity changes. In particular, the penalty for the yaw angle is higher for higher velocities and lower for lower velocities. The penalty for the lateral position error exhibits the opposite trend.
    }
    \label{fig:oc}
\end{figure}

Fig.~\ref{fig:oc} shows trajectories of the closed-loop simulations of the lane-change optimal controller with the cost function in \eqref{eq:lane_oc} in CarSim. At time 0~s, all control parameters are set to zero, $\theta_0=0$. The Kalman filter quickly notices that the cost function parameters need to be positive in order to execute the lane change. This is because of the model-based structure of the calibration module that uses a kinematic bicycle model. After three lane-change maneuvers (around 20~s), the Kalman filter has found a set of parameters that works well---as defined by the specifications---given the vehicle velocity of around 70~km/h. Then, the velocity is decreased at 50~s. As a result, the Kalman filter adapts the cost function parameters in order to optimize the specifications given the altered vehicle velocity.

\subsection{Results for State Feedback Controller}
\label{ssec:sf_results}
Second, we analyze a lane-change controller with the control law
\begin{align}
\label{eq:lane_fsf} 
    \dot \delta_k = 
    \begin{bmatrix}
        \theta_y & \theta_\psi & \theta_\delta
    \end{bmatrix}
    \begin{bmatrix}
        p_{Y,k}-p_{Y,k,{\rm ref}}\\ \psi_k \\ \delta_k
    \end{bmatrix}.
\end{align} 

\begin{figure}
\centering 
\begin{tabular}{ll} 
\begin{tikzpicture}[every plot/.append style={line width=0.5mm},every mark/.append style={solid},]
\begin{axis}[
    xmin=0, xmax=120,
    ymin=-3.5, ymax=3.5,   
   ylabel={Lateral pos. [m]},  
   ytick = {-1.5,0,1.5}, 
   xticklabels={,,,,,,,,,,,,,,,}, 
    legend style={at={(0.5,1.02)},anchor=south},  
    ymajorgrids=true,
    legend columns=4,
    transpose legend,
    grid style=dashed,
    width=8cm, 
    height=4.2cm,
    mark repeat=5,
    y label style={at={(axis description cs:0.05,.5)},rotate=0,anchor=south},
    legend cell align={left}
	] 
]	  
\addplot[color=black] table[x index=0,y index=1] {data_CarSim/fsf_pY.txt};   
  
\end{axis}
\end{tikzpicture} 
\\
\begin{tikzpicture}[every plot/.append style={line width=0.5mm},every mark/.append style={solid},]
\begin{axis}[
    xmin=0, xmax=120,
    ymin=30, ymax=80,   
   ylabel={Velocity [km/h]},  
  % yticks = {5,10}, 
   %yticklabels={ },  
   xticklabels={,,,,,,,,,,,,,,,}, 
    legend style={at={(0.5,1.02)},anchor=south},  
    ymajorgrids=true,
    legend columns=4,
    transpose legend,
    grid style=dashed,
    width=8cm, 
    height=4cm,
    mark repeat=5,
    y label style={at={(axis description cs:0.05,.5)},rotate=0,anchor=south},
    legend cell align={left}
	] 
]	     
\addplot[color=black] table[x index=0,y index=1] {data_CarSim/fsf_vel.txt}; 
  
\end{axis}
\end{tikzpicture} 
\\
\begin{tikzpicture}[every plot/.append style={line width=0.5mm},every mark/.append style={solid},]
\begin{axis}[
    xmin=0, xmax=120,
    ymin=-7.2, ymax=0.2,   
   ylabel={Control Param.}, 
   xlabel={Time [s]}, 
  % yticks = {5,10}, 
   %yticklabels={ },  
   %xticklabels={,,,,,,,,,,,,,,,}, 
    legend style={at={(1,0)},anchor=south east},  
    ymajorgrids=true,
    legend columns=3, 
    grid style=dashed, 
    width=8cm, 
    height=4.2cm,
    mark repeat=5,
    y label style={at={(axis description cs:0.045,.5)},rotate=0,anchor=south},
    legend cell align={left}
	] 
]	     
\addplot[color=black] table[x index=0,y index=1] {data_CarSim/fsf_theta.txt}; 
\addplot[color=mst2,densely dashed] table[x index=0,y index=2] {data_CarSim/fsf_theta.txt}; 
\addplot[color=mst3,densely dashdotted] table[x index=0,y index=3] {data_CarSim/fsf_theta.txt};

\legend{$10\cdot \theta_{y}$,$\theta_{\psi}$,$\theta_{\delta}$};
 
\end{axis}
\end{tikzpicture} 
\end{tabular}
    \caption{Calibration of state feedback controller. 
    }
    \label{fig:fsf}
\end{figure}
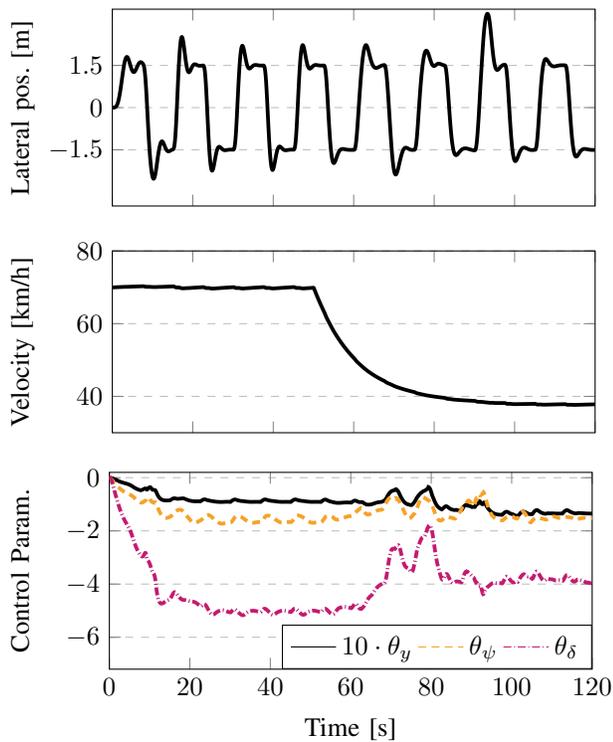

Fig.~\ref{fig:fsf} shows trajectories of the lane-change state feedback controller \eqref{eq:lane_fsf} with the CarSim simulator. At time 0~s, all control parameters are set to zero, $\theta_0=0$. Then, the Kalman filter calibration module realizes that the gains need to be negative in order to execute the lane change. Here, too, the control parameters are adapted in the ``right" direction due to the model-based structure that uses a kinematic bicycle model. While the first lane-change maneuver is executed successfully, the performance is not ideal yet, cf., lateral position oscillations around 0--8~s. Consequently, the Kalman filter continues to calibrate the control parameters to further improve upon the system performance (measured with respect to the specifications). After 25~s the control parameters enter a repetitive pattern.  
At time 50~s, the velocity is decreased. As a result, the control parameters are adapted by the Kalman filter in order to perform the lane changes at lower velocity. For example, $\theta_\delta$ is decreased and $\theta_y$ is increased. Note that the different scales of the control parameters are partially caused by different units, e.g, the lateral position error in meters takes higher values than the angles in radians.

\subsection{Extended Discussion}
Due to the model mismatch between the kinematic single track vehicle model used in the feedback controller and/or the Kalman filter-based adaptation module and the high-fidelity vehicle simulator in CarSim, the estimated control parameters are expected not to converge. 
Instead, the control parameters can enter a limit cycle, which is best observed in Fig.~\ref{fig:oc}, e.g., between 30s--50s and between 100s--120s.
The model mismatch is important to calibrate the controller to achieve the best closed-loop performance.
This is a key advantage of the proposed method, because it leverages the fact that the feedback controller uses a simplified control-oriented model in order to control a complex dynamical system. 
Note that some industrial applications may require non-adaptive control parameters, e.g., due to regulations.
In such a scenario, the calibration may be performed at production stage by freezing the control parameters at the time that the limit cycle is entered, or by averaging the control parameters over one limit cycle. 
Furthermore, state-dependent control parameters can still be considered by interpolating between pre-computed sets of control parameters, which is similar to gain scheduling.

For the online implementation of the proposed method, the Kalman filter uses a sliding window of length $N$ in order to calibrate the control parameters, $\theta$.
While this is different from the standard Kalman filter implementation, this sliding horizon is important to be captured for calibrating control parameters in order to evaluate the evolution of the dynamical system under a given control parameter realization. 
This setup is comparable to moving horizon estimation~\cite{rawlings2006particle}, where a sliding window is needed in order to capture the behavior of a dynamical system under constraints.

\subsection{Computation Times}
We implemented the proposed adaptation method on a dSPACE MicroAutoBox-II rapid prototyping unit, which is equipped with a 900 MHz PowerPC real-time processor (IBM PPC 750GL) and 16 MB of RAM.  
The dSPACE MicroAutoBox-II rapid prototyping unit reflects the current and next-generation capabilities of embedded micro-controllers in automotive systems. We utilized an implementation in Simulink using the Matlab function block with automatic C code generation. 
 
For the optimal control CarSim example in Section~\ref{ssec:oc_results}, the maximum computation time on the MicroAutoBox-II was 0.0196~s.
For the state feedback controller example in Section~\ref{ssec:sf_results}, the maximum computation time on the MicroAutoBox-II was 0.0205~s.
For both examples, the computation times are well within the sampling period of 0.1~s used in this implementation, thus validating the real-time capabilities of the proposed method in automotive embedded platforms.

%%%%%%%%%%%%%%%%%%%%%%%%%%%%%%%%%%%%%%%
%%%%%%%%%%%%%%%%%%%%%%%%%%%%%%%%%%%%%%%
%%%%%%%%%%%%%%%%%%%%%%%%%%%%%%%%%%%%%%%
%% SECTION
%%%%%%%%%%%%%%%%%%%%%%%%%%%%%%%%%%%%%%%
%%%%%%%%%%%%%%%%%%%%%%%%%%%%%%%%%%%%%%%
%%%%%%%%%%%%%%%%%%%%%%%%%%%%%%%%%%%%%%%
\section{Conclusion}
\label{sec:conclusion}
This paper proposed a method to calibrate controller parameters. The method is implemented in a recursive fashion using a Kalman filter that estimates control parameters rather than the system's state. The calibration is driven by a training objective, which encompasses specifications on the operation of the closed-loop system. 
The main benefits of the proposed method are the low computational requirements, low data storage requirements, and a relatively high flexibility e.g, in embedding non-differentiable objectives in the control system.
Simulation results showed that the method is able to calibrate parameters for a wide range of controllers, such as state feedback, optimal, PID, $H_{\infty}$, sliding mode, dynamic output feedback, and neural network controllers. 
In particular, the method was shown to learn the control parameters quickly and robustly (approximately 24\% average decay factor of closed-loop cost). 
Further, the method was shown to adjust the parameters to systematic disturbances, thereby effectively reducing closed-loop cost of the system operation (approximately 29\% improvement on tracking precision).
Further, a simulation study with the high-fidelity vehicle simulator CarSim showed that the method can calibrate controllers of a complex dynamical system online, which suggests its appropriateness to operate on real-world systems.
Finally, we showed that the algorithm is implementable on current and next-generation embedded platforms for automotive applications using a dSPACE MicroAutoBox-II rapid prototyping unit.
Overall, this paper verifies the algorithm on a vehicle model that is close to a physical vehicle and on an embedded platform suitable for automotive applications.
Hence, the findings in this paper also validate the applicability of the proposed algorithm today's and near future vehicles.

\renewcommand{\arraystretch}{1}

\section*{Appendix: CarSim Parameters}
Table~\ref{tab:carsim} shows the main modules and vehicle parameters used for the CarSim simulation. 
\begin{table}[t]
    \centering
    \caption{
    CarSim Vehicle and Simulation Parameters
    }
    \begin{tabular}{l|llll} 
        \toprule
        \multicolumn{2}{l}{Vehicle (Rigid sprung mass): B-Class, Hatchback w/Hybrid Powertrain}\\ 
        $\sbullet$ Mass & 1134~kg  
        \\
        $\sbullet$ Wheel base & 2.6~m  
        \\
        $\sbullet$ Roll/Pitch/Yaw inertia & 440.6~/~1343.1~/~1343.1~kg m\textsuperscript{2} 
        \\ 
        \midrule
        \multicolumn{2}{l}{Aerodynamics: B-Class, Hatchback Aero}
        \\ 
        $\sbullet$ Frontal area & 1.6~m2
        \\ 
        \midrule 
        \multicolumn{2}{l}{Suspension (Front): B-Class, Hatchback - Front}  
        \\
        $\sbullet$ Unsteered/Steered unsprung mass & 10.31~/~25.39~kg (each side) 
        \\
        $\sbullet$ Wheel center distance & 1.48~m
        \\  
        \midrule
        \multicolumn{2}{l}{Suspension (Rear): B-Class, Hatchback - Rear}    
        \\
        $\sbullet$ Unsteered/Steered unsprung mass & 27.25~/~0~kg (each side) \\
        $\sbullet$ Wheel center distance & 1.485~m
        \\  
        \midrule
        \multicolumn{2}{l}{
        Springs, Dampers, and Compliance: B-Class, Hatchback - Front}
        \\ 
        $\sbullet$ Spring rate & 28~N/mm
        \\ 
        $\sbullet$ Spring friction & 20~N
        \\  
        \midrule
        \multicolumn{2}{l}{
        Springs, Dampers, and Compliance: B-Class, Hatchback - Rear}
        \\ 
        $\sbullet$ Spring rate & 35~N/mm
        \\ 
        $\sbullet$ Spring friction & 20~N 
        \\  
        \midrule
        \multicolumn{2}{l}{
        Tires: 185/65 R15 }
        \\
        $\sbullet$ Mass & 16~kg (each)
        \\
        $\sbullet$ Spin/Yaw/Roll Inertia & 0.8~/~0.4~/~0.4~kg m\textsuperscript{2} (each)
        \\
        $\sbullet$ Rolling radius & 287~mm
        \\  
        $\sbullet$ Rolling resistance & 0.008
        \\
        \midrule
        \multicolumn{2}{l}{
        Powertrain: 20~kW x2, Electric AWD Hi-Pw, 3.905 Ratio}
        \\ 
        \midrule
        \multicolumn{2}{l}{
        Brake System: (Hbk): MC Press, No ABS, No Rr  }
        \\ 
        \midrule
        \multicolumn{2}{l}{
        Steering System: B-Class, Hbk: Power, R\&P}  
        \\  
        \midrule
        \multicolumn{2}{l}{
        Road: Straight Lane }
        \\
        $\sbullet$ Friction & $\mu=0.9$
        \\
        \midrule
        \multicolumn{2}{l}{
        Simulation Parameters: ode45 solver}
        \\
        $\sbullet$ Simulation sampling time & 0.5~ms
        \\
        $\sbullet$ Controller sampling time & 0.1~s
        \\
        \bottomrule
    \end{tabular}
    \label{tab:carsim}
\end{table}

\bibliography{bib/mpc_cal}
\bibliographystyle{ieeetr}

\end{document}